\documentclass[manuscript]{acmart}

\AtBeginDocument{%
  }

\setcopyright{acmlicensed}
\copyrightyear{2025}
\acmYear{2025}
\acmDOI{XXXXXXX.XXXXXXX}

\acmConference[Conference acronym 'XX]{Make sure to enter the correct
  conference title from your rights confirmation emai}{June 03--05,
  2018}{Woodstock, NY}
\acmISBN{978-1-4503-XXXX-X/18/06}

\usepackage{todonotes}
\usepackage{enumitem}
\usepackage{xspace}
\usepackage{listings}

\begin{document}
\newcommand{\STATICFULL}{Static Prompt Refinement Control\xspace}
\newcommand{\DYNAMICFULL}{Dynamic Prompt Refinement Control\xspace}
\newcommand{\STATIC}{Static PRC\xspace}
\newcommand{\DYNAMIC}{Dynamic PRC\xspace}
\newcommand{\OPTIONMODULE}{Option Module\xspace}
\newcommand{\CHATMODULE}{Chat Module\xspace}

\title{Dynamic Prompt Middleware: Contextual Prompt Refinement Controls for Comprehension Tasks}
\author{Ian Drosos}
\email{t-iandrosos@microsoft.com}
\author{Jack Williams}
\email{jack.williams@microsoft.com}
\author{Advait Sarkar}
\email{advait@microsoft.com}
\author{Nicholas Wilson}
\email{nicholas.wilson@microsoft.com}
\affiliation{%
  \institution{Microsoft Research}
  \city{Cambridge}
  \country{UK}
}


\begin{abstract}
Effective prompting of generative AI is challenging for many users, particularly in expressing context for comprehension tasks such as explaining spreadsheet formulas, Python code, and text passages. 
Prompt middleware aims to address this barrier by assisting in prompt construction, but barriers remain for users in expressing adequate control so that they can receive AI-responses that match their preferences.

We conduct a formative survey ($n=38$) investigating user needs for control over AI-generated explanations in comprehension tasks, which uncovers a trade-off between standardized but predictable support for prompting, and adaptive but unpredictable support tailored to the user and task.
To explore this trade-off, we implement two prompt middleware approaches: \DYNAMICFULL (\DYNAMIC) and \STATICFULL (\STATIC).
The \DYNAMIC approach generates context-specific UI elements that provide prompt refinements based on the user's prompt and user needs from the AI, while the \STATIC approach offers a preset list of generally applicable refinements.

We evaluate these two approaches with a controlled user study ($n=16$) to assess the impact of these approaches on user control of AI responses for crafting better explanations. 
Results show a preference for the \DYNAMIC approach as it afforded more control, lowered barriers to providing context, and encouraged exploration and reflection of the tasks, but that reasoning about the effects of different generated controls on the final output remains challenging.
Drawing on participant feedback, we discuss design implications for future \DYNAMIC systems that enhance user control of AI responses.
Our findings suggest that dynamic prompt middleware can improve the user experience of generative AI workflows by affording greater control and guide users to a better AI response. 
\end{abstract}



\begin{teaserfigure}
  \includegraphics[width=.95\textwidth]{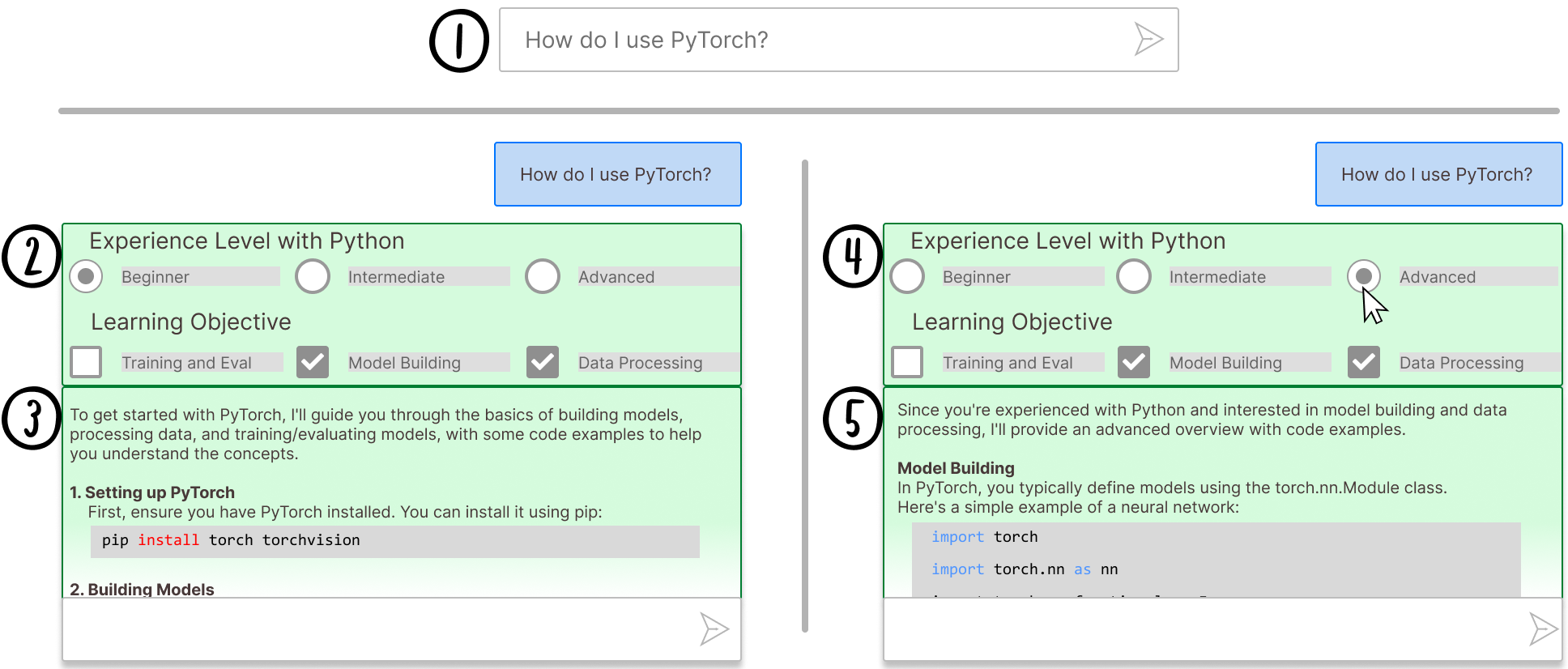}
  \caption{Schematic overview of Dynamic Prompt Refinement Control interface for generating inline prompt refinement options to increase user control of AI-generated explanations (derived from real system use). 
  (1) User prompts the system.
  (2) The \OPTIONMODULE takes the user's prompt as context to generate prompt options which provide prompt refinements for the user to select. 
  (3) The \CHATMODULE uses the user's prompt and pre-selected options to generate an initial response.
  (4) The user can initiate refinements by selecting their preferred options in the UI.
  (5) On each change, the \CHATMODULE regenerates the response based on the new selections.}
  \Description{}
  \label{fig:teaser}
\end{teaserfigure}

\received{10 October 2024}

\maketitle

\section{Introduction}
\label{sec:Introduction}



As Generative AI (GenAI) tools achieve widespread penetration in knowledge workflows, an increasingly common observation made by user studies is that \emph{prompting}, i.e., the skillful and effective use of natural language for steering and eliciting GenAI behavior, is challenging for many users in many contexts \cite{zamfirescu2023johnny,tankelevitch2024metacognitive,drosos2024duck,liu2023gam}. The challenges range from the increased explicit metacognitive demands of prompting, to the burden of explaining detailed requirements and context in textual format, to the opaque and unpredictable mapping of naturalistic language queries to system behavior.

In response, researchers have noted the limitations of the chat paradigm as an interface for GenAI tools, suggesting that we go ``beyond chat'' \cite{ma2024exploreLLM}, and numerous scaffolded writing tools that integrate direct manipulation with text have been developed \cite{buschek2024collage}. In particular, the use of graphical user interface elements known as ``prompt middleware'' \cite{macneil2023promptmiddleware} can help users rapidly construct and compose more elaborate prompts. We survey the prior work in this space in Section~\ref{sec:RelatedWork}. However, a challenge with scaling the prompt middleware approach is that to develop effective middleware requires an understanding of the user, their task, and effective prompting approaches. For effective prompt middleware, developers need to carefully characterize tasks, users, and prompting strategies.

\textbf{Our approach.} In this paper, we propose instead to \emph{generate} prompt middleware customized to the user and task. Our approach (Figure~\ref{fig:teaser}) is to instantiate a language model agent (the ``\OPTIONMODULE'') whose responsibility is to analyze the user's prompt and generate a set of options for refining that prompt. These options are rendered using graphical elements such as radio buttons, checkboxes, and free text boxes, which the user can select and edit to cause deterministic additions, removals, and modifications of their original prompt, which is passed as context to a conventional language model agent (the ``\CHATMODULE'') which fulfills the request. Our implementation is detailed in Section~\ref{sec:DesignImpl}.

To our knowledge, such \emph{dynamic prompt middleware} has not been well studied in general applications of GenAI, although there is precedent work in the specific domain of code generation \cite{cheng2024biscuit}, and there is a long history of dynamic interface generation, particularly in data visualization (Section~\ref{sec:RelatedWork}). As a novel test bed for dynamically generated controls, we select comprehension and learning tasks. In particular, our study tasks (Section~\ref{sec:Methods:Tasks}) focus on the use cases of using GenAI to explain spreadsheet formulas, python code, short text passages, and as a teaching aid for data analysis and visualization concepts. These are, broadly speaking, sensemaking tasks \cite{pirolli2005sensemaking}. According to a recent taxonomy of GenAI use developed by \citet{brachman2024knowledge}, they fall under the category of ``advice'' tasks, distinguished from ``creation'' tasks (using GenAI to produce an artifact for direct or indirect use in a workflow), ``information'' tasks (using GenAI to retrieve facts from a database or the web), and ``automation'' tasks (using GenAI to control software and automate other computer actions).

We are interested in comprehension and learning tasks for three reasons. First, they are a common and important use case, which increases the ecological validity of our work. Second, because there is a great diversity among tasks and users that results in a need for personalization and fine-grained control (``out of the box'' GenAI responses to a user's initial prompt often do not satisfy the user's comprehension requirement, as we find through our formative study and which has also been reported in previous work (e.g., \cite{drosos2024duck})). And finally, because their primary objective is to further the quality of human thought and understanding. As recent work has noted, the introduction of GenAI into knowledge workflows can cause the deterioration of human thinking, and researchers have called for the critical design development of GenAI tools to reverse this trend and even enable new ``Tools for Thought'' \cite{sarkar2024aiprovocateur, sarkar2024copilot,sarkar2024intention,MicrosoftT4T}.

We make the following contributions:
\begin{itemize}
    \item We report a formative survey ($n=38$) including an interactive design probe to understand user needs around customized explanations of GenAI output (Section~\ref{sec:Formative}). We find that users desire greater direct control over AI responses than afforded by traditional chat interfaces, and that controls should adapt to the task and user.
    \item We develop two variations of an interface with graphical controls for steering AI responses, one with static controls, and one with dynamic controls that are generated based on the user's initial prompt (Section~\ref{sec:DesignImpl}). We report a controlled, within-subjects comparison of the two interfaces ($n=16$, Section~\ref{sec:Methods}), finding that users preferred dynamic controls because they lower barriers to providing context, offer guidance for better prompts, facilitate greater exploration and reflection, but are viewed as more complex and more difficult to reason about in terms of their effects (Section~\ref{sec:Results}).
    \item We derive design implications for dynamic controls, such as greater control of how the AI interprets and applies options, leveraging of user context and data to generate helpful controls, expanding dynamic prompt middleware to richer modalities, and discuss our findings in connection with related work (Section~\ref{sec:Discussion}).
\end{itemize}

\section{Related Work}
\label{sec:RelatedWork}
\paragraph{Challenges to effective prompting and steering GenAI}
Previous work has identified several challenges for end users in writing instructional prompts for GenAI \cite{zamfirescu2023johnny}, such as finding appropriate levels of vocabulary and grammar that correspond to the level of abstraction of the system output (known as the ``fuzzy abstraction matching'' problem) \cite{liu2023gam}, the burden of elaborating the task context needed for the GenAI system to produce relevant and useful responses \cite{drosos2024duck}, or of having sufficient metacognitive awareness to understand how to apply one's knowledge of the domain and prompting to steer and verify GenAI output \cite{tankelevitch2024metacognitive}.

A related issue that commonly manifests as a prompting challenge is the difficulty in decomposition of tasks and fine-grained steering and control of each sub-task. 
Research has explored various approaches to solving this problem, such as rendering each individual sub-task as interactive elements prior to generation \cite{ma2024exploreLLM,kazemitabaar2024improving} and rendering the generated output as a series of editable steps \cite{liu2023gam,ferdowsi2023coldeco}.

\paragraph{``Beyond chat'' interfaces for GenAI and automatically generated interfaces}
Another issue is the cumbersome and inconsistent nature of text entry as an interaction mechanism. 
To remedy this, researchers have proposed various strategies for going beyond chat as the sole interaction metaphor, and blending text with graphical user interface elements. 
The automatic generation of custom user interfaces has long been studied \cite{gajos2004supple,gajos2007automatically,nichols2003personal,nichols2006uniform,nichols2006huddle,vaithilingam2019bespoke,chen2022pi2,zhang2018precision}, but GenAI poses new problems and new opportunities \cite{jovanovic2024generative,moran2024ui}. 
For instance, researchers have proposed generating relevant customization widgets for data visualizations on an ad-hoc basis \cite{Vaithilingam2024DynaVis, dibia-2023-lida,chen2022nl2interface}. 
Another approach is to generate user interface elements that help the user refine specific aspects of the generated artifact (e.g., code) \cite{cheng2024biscuit}. 
The latter is an example of a general category of techniques that use graphical interface elements for manipulating prompts, sometimes referred to as ``prompt middleware'' \cite{macneil2023promptmiddleware}, to ease the cognitive burdens of text entry and manipulation. 

Researchers have also proposed ideation tools that help users develop queries based on their own document corpus \cite{dhole2023interactive}, to scaffold the metacognitive process of reasoning about one's own goals.
Researchers have also developed ways to help users manage the large volume of information generated by language models, such as by rendering them in a hierarchy of abstraction levels \cite{suh2023sensecape}. 
More generally, there has been much experimentation with the integration of graphical user interface elements with large language models, particularly in reading and writing tools, which are surveyed by \citet{buschek2024collage}, who describes it as ``collage''.

\paragraph{A note on the term ``explanation''} 
In the context of AI, the term explanation is overloaded and most commonly refers to a line of research known as explainable AI (XAI) \cite{gunning2019xai,xu2019explainable,dwivedi2023explainable,hoffman2018metrics,goebel2018explainable,holzinger2022explainable}, which typically focuses on explaining aspects of the model, its training data, its inference process, etc. 
The typical use of explanations is as a decision support mechanism, to help the user evaluate and trust (or not trust) the AI output and decide what next interaction to perform with the AI system. 
However, we do not use the term in this sense; by explanation we instead mean aids for comprehension more generally. 
In our study tasks, we use AI to help ``explain'' code or text -- these explanations are not about the mechanism of the AI system. 
As much previous work has noted, the explainability concerns of end-users are much broader than just the AI system itself \cite{miller2019explanation,ehsan2024explainability,ehsan2021explainable,ehsan2021operationalizing,ehsan2022human,ehsan2019automated,ehsan2020human,ehsan2021expanding,ehsan2023charting,ehsan2024seamful,sarkar2024llmscannotexplain,sarkar2022explainable}, it is this broader sense in which we use the term. 
Recent work on code explanations has found user needs around these kinds of explanations, which resulted in the design implications that explanations be \textit{adaptable} and \textit{controllable} \cite{YanIvie2024}, which we also find in our formative survey and use as driving design goals for the systems presented in the paper.

\paragraph{In summary,} previous work has identified serious challenges faced by users in reasoning about one's own goals while prompting, and then rapidly customizing prompts to effectively steer the output of Generative AI. While previous work has explored ways of scaffolding users to improve their metacognitive and prompting capabilities, and developed ``prompt middleware'' to lessen the burden of text entry and manipulation, close attention has not yet been paid to generating middleware dynamically, adapted to the user's specific task at hand, in the context of comprehension and learning tasks. As previously argued, these are a common and diverse set of GenAI tasks for which contextualized prompt middleware may prove useful; it is this idea that we develop and evaluate in this paper.




\section{Formative Survey and Design Probe}
\label{sec:Formative}
We conducted a formative survey of employees with access to GenAI tools, at a large software company. 
Our survey focused on how explanations generated for data-driven tasks within spreadsheets might be improved by showing two scenarios to participants and asking several questions around the effectiveness of the explanations that GenAI gave, strategies on verifying and correcting GenAI output, and questions around the importance of learning the concepts the AI used to complete a task. 

The survey also contained a design probe of a preliminary design to gather design goals for systems that provide greater control of AI responses. We collected participant ratings of the importance of having such control in GenAI interactions, the characteristics of AI responses that participants needed control over, and feedback on how to improve the design to better control AI explanations, which we describe below.
For brevity we focus only on results that directly informed the design of our two approaches (\DYNAMICFULL and \STATICFULL) developed for the study.


\subsection{Survey setup}
We sent a survey to employees of a large software company looking for participants that had previous experience using GenAI tools like Copilot~\cite{CopilotSite}. 38 participants (F1-F38) responded to the survey. We collected participant demographic information using a previously developed questionnaire assessing GenAI experience~\cite{drosos2024duck}. The majority of participants stated they ``Regularly use one or more'' GenAI tools (25 or 65\%), with 11 saying they ``Occasionally use'' (29\%), 1 ``Casually tried'' (3\%), and 1 ``had not tried'' (3\%) GenAI tools. 
To improve the validity of responses, participants were asked to pretend they were part of a detailed scenario to understand the AI-generated explanations so that they could explain them to others.

\subsection{Design Probe}
\begin{figure}
    \centering
  \includegraphics[width=0.8\textwidth]{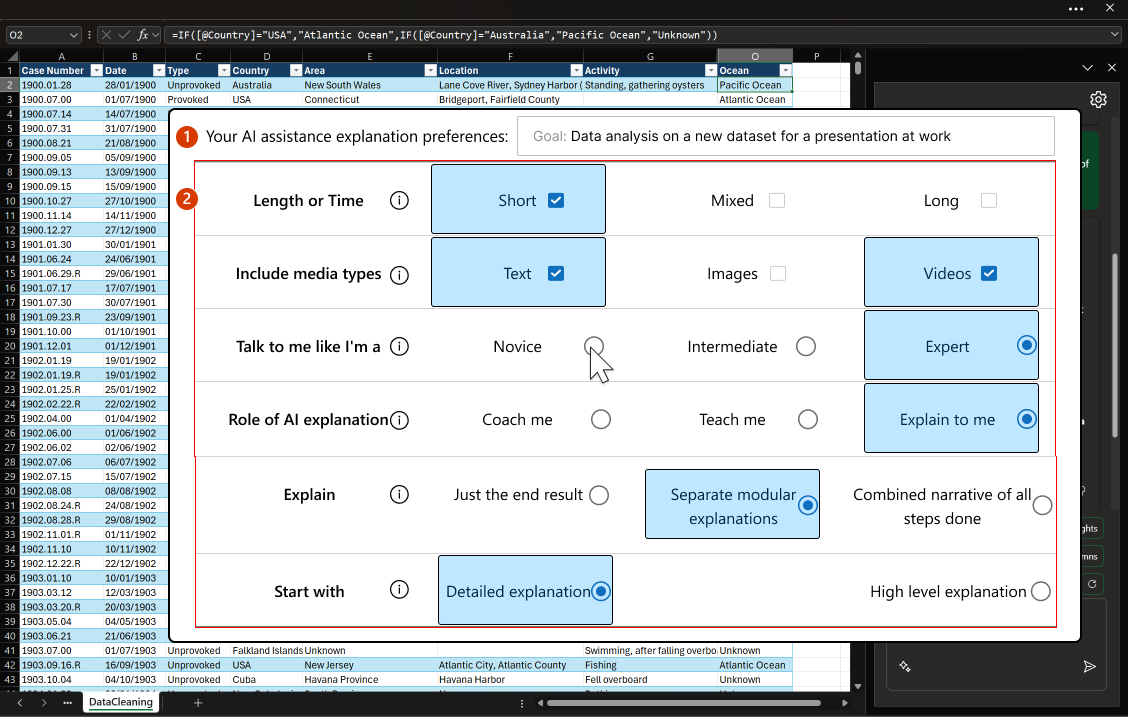}
  \caption{One frame of the control interface presented to users as part of the formative design probe to collect feedback for a final design. (1) Shows an open ended text-box that users can input their goal which prompts the AI to select relevant options. (2) A grid of option elements that users can select which would adapt the AI's responses.}
  \Description{}
  \label{fig:DesignProbe}
\end{figure}

Participants were shown a design of a potential interface for providing control over AI responses (Figure~\ref{fig:DesignProbe}), walked through several steps of a scenario of the designs usage which explained the various features, and asked several questions around their needs for control while interacting with GenAI tools. 
This design contained a grid of hand-crafted options that were potentially useful to select from when crafting AI responses (Figure~\ref{fig:DesignProbe}.2). 
The design also contained a text-box where the user could give a prompt to the system that reflected their goal for their interaction with the AI, which would help select potentially useful selections (Figure~\ref{fig:DesignProbe}.1). 
However, the options themselves (e.g., Length or Time, Start with, etc.) were static and could not be changed by the user. 

The goal of this design was to elicit other types of options that users might find useful for controlling AI responses, which might then inform a sort of ``prompt-option toolbox'' to quickly control various elements of the AI response. 
We also elicited other improvements and features that could assist in user control over AI responses from our participants.   

\subsection{Formative Study Results: Participant Preferences for Control Over AI Explanations}
\label{sec:Formative:Results}
Participants agreed that having control over the AI response (like what was provided in the design probe) was important to them (Median=4.0 [Agree], Mean=4.3).
Participants believed control was important for their own understanding of what the AI had done (F2), help the AI \emph{``adapt to the user in the context of their daily work''} (F3), and helpful for getting better responses from the AI (F14). 
Participants wanted the AI to respond in a predictable manner (F21), and saw the control afforded by the design as a way to \emph{``tailor the experience to your specific needs''} (F22, F15-17) and put users in control of the AI (F28). This feedback led to our first design goal:

\begin{quote}
\textbf{D.1} Generative AI tools should afford users direct control of the AI's responses.
\end{quote}

Participants also wanted options to be generated based on the prompt they gave it (F7), quick to change within the chat interface (F10), and adapted to the type of task the user was trying to accomplish with the AI (F3, F4, F8, F12, F25, F30, F32, F36). Participants needed flexibility in the types of responses they received from AI (F19) and did not want to have to \emph{``do minor adjustments to the prompt manually''} (F21). However, participants also wanted the design to go a step further by allowing custom options based on their needs (F38). This desire for control to be flexible to the user and their task led to our second design goal:

\begin{quote}
\textbf{D.2} Control affordances should be flexible, so users can quickly correct AI assumptions, and \emph{dynamic}, so control surfaces adapt to the task and the user.
\end{quote}

We note that both design goals align with two of the design goals from the results of recent related research on AI generated explanations, namely that they should be \emph{``controllable''} (D.1) and \emph{``adaptable''} (D.2) \cite{YanIvie2024}.

While participants were generally happy with the affordances and options provided in the design, we received feedback which informed several improvements for the design and implementation of the \DYNAMICFULL system (\DYNAMIC, detailed in Section~\ref{sec:DesignImpl}) and the \STATICFULL system (\STATIC, detailed in Section~\ref{sec:DesignImpl:Static}).
In general, participants wanted to save their options for follow-up prompts and sessions (F1, F3, F31, F33), modify options dynamically (F5), and be provided descriptions of the options to understand what they might do to a response (F2, F10). We also collected what type of options participants would like to see via an open-text question: ``What characteristics of AI explanations it important for you to have control over?''. Participant responses to this question informed the options available in the \STATIC system (Section~\ref{sec:DesignImpl:Static}). 


\section{Tool Design and Implementation}
\label{sec:DesignImpl}
\begin{figure}
    \centering
  \includegraphics[width=1.0\textwidth]{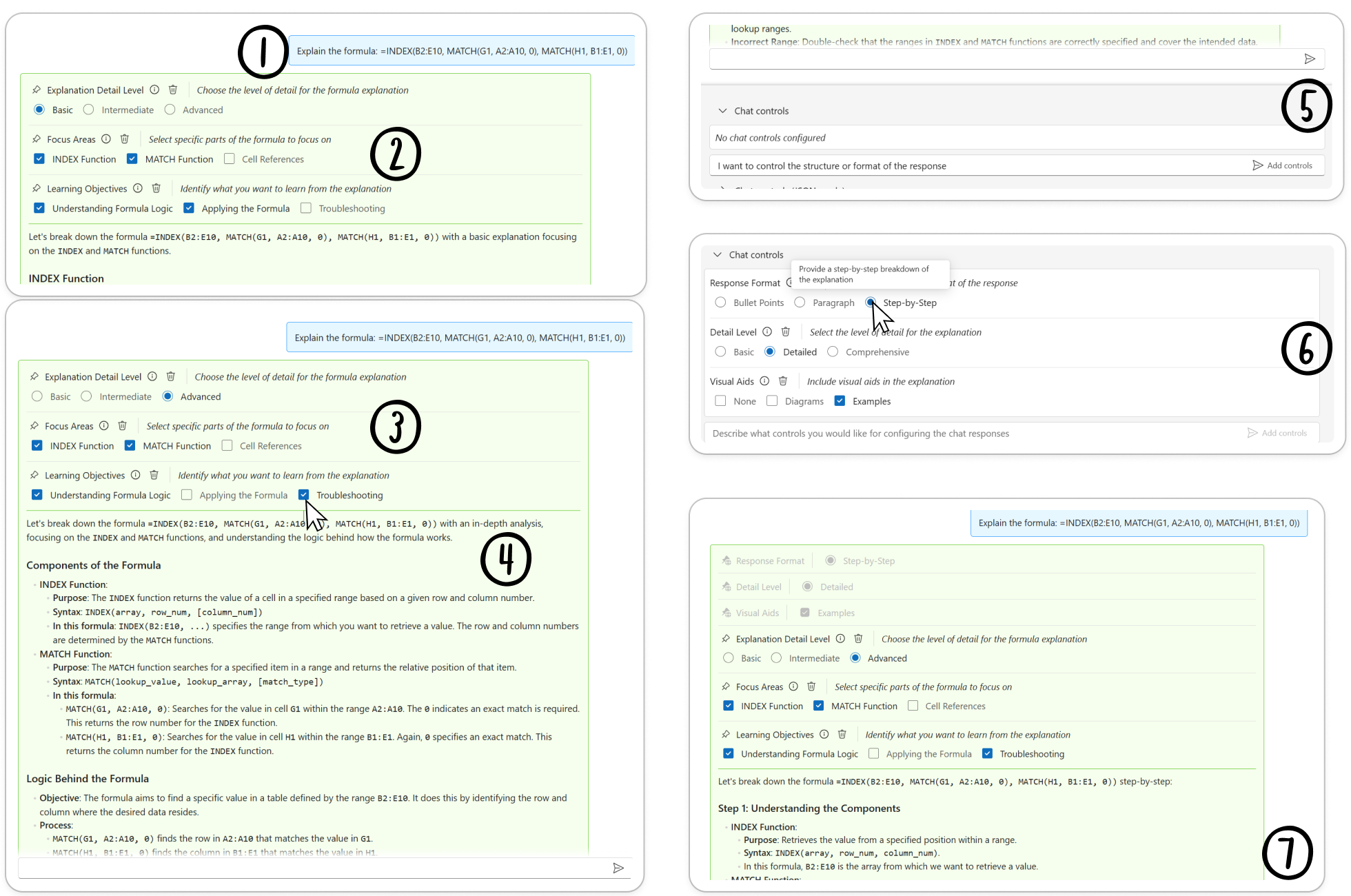}
  \caption{User flow with the \DYNAMIC system. (1) User submits their prompt. (2) The \OPTIONMODULE generates a set of options to help steer the \CHATMODULE's response. (3) User can update the refinements sent to the \CHATMODULE by clicking their preferences. (4) On change, \CHATMODULE regenerates the response with the new chosen refinements. (5) User can request controls through NL prompting. (6) The \OPTIONMODULE generates a set of session options based on this prompt. (7) The session options apply to the current and every subsequent response from the \CHATMODULE.}
  \Description{}
  \label{fig:userFlow}
\end{figure}

We now describe the design and implementation of our dynamic prompt middleware system, \DYNAMIC, and detail how we applied the design goals informed by our formative survey. As part of this work, we designed a second system to compare, \STATIC. The \STATIC system is a simplified instance of the \DYNAMIC system, therefore we focus our description on the latter. Both systems are web-based, built using React \cite{react} and TypeScript \cite{typescript}. All LLM completions are generated using \texttt{gpt4-turbo}.

\subsection{Using dynamic prompt middleware}
We describe a user flow within our system, illustrated in Figure~\ref{fig:userFlow}. A user wants to generate an AI explanation of their spreadsheet formula, and enters the prompt ``Explain the formula:  
\texttt{=INDEX(B2:E10, MATCH(G1, A2:A10, 0), MATCH(H1, B1:E1, 0))}'' (1). 
The \DYNAMIC system uses this input, along with any previous conversation history, to generate prompt options for the input. Given the context, the system generates three separate \emph{inline options}: ``Explanation Detail Level'', ``Focus Areas'', and ``Learning Objectives'' with several choices for each (2). 
The system always selects initial values for each option, the prompt refinements. The user inspects each option and sees that they would like a more advanced explanation that explains how to fix an issue they were having with the formula. So, the user changes Explanation Detail Level from Basic to Advanced, and changes the Learning Objective from Applying the Formula to Troubleshooting (3). 
As the user modifies the selection, the chat response updates automatically using the new prompt refinements (4). 
The user inspects the response and realizes they want a different structure to the AI explanation of the formula. 
The user goes to the Chat control panel and in the open text field types ``I want to control the structure or format of the response'' (5). 
Based on the user's request, the system generates \emph{session options}, including for the ``Response Format'' (which includes Bullet Points, Paragraph, and Step-by-Step). 
The user selects Step-by-Step which organizes the explanation into discrete steps which the user finds preferable to the previous response (6). 
Using the newly added session options, the \CHATMODULE regenerates the response (7).

\subsection{Two-tier \DYNAMICFULL}
\label{sec:DesignImpl:Dynamic}
We developed the \DYNAMIC system to automatically generate prompt options and refinements for an LLM interface (D.2). The options generated by the \DYNAMIC system are streamed incrementally and presented to the user to afford greater control of the AI response to the prompt (D.1). The user can then select from these options, modify existing options, generate their own options via a natural language prompt, or even define prompt options directly using JSON.  

We designed and implemented a \emph{two-tier} system that uses two sets of prompt options. The first supports inline response-level dynamic controls generated for every user input, and the second supports session-level prompt controls that apply throughout the entire session with the AI.
This two-tier system provides users the flexibility between control (providing options for the entire session) and dynamism (system generation of controls adapted to each prompt for the user to explore).

\subsubsection{Prompt options}
\begin{figure}
    \centering
  \includegraphics[width=0.8\textwidth]{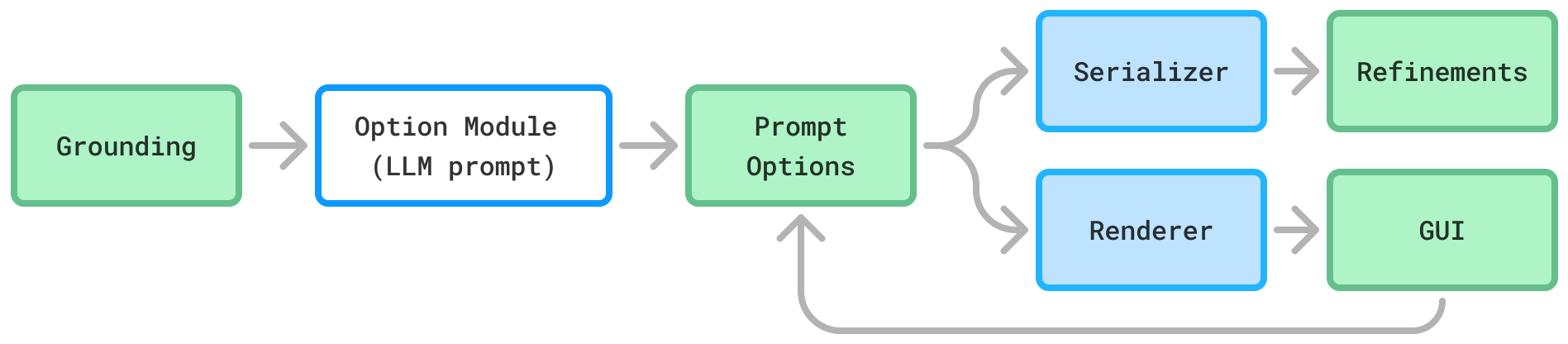}
  \caption{Prompt options model}
  \Description{}
  \label{fig:prompt_options}
\end{figure}
The central component to our approach is \emph{prompt options} which is used to drive prompt refinements. Prompt options are represented as a JSON data structure that describes the options and their current value; additionally, prompt options have a renderer and serializer. The renderer takes a set of prompt options and displays them as a GUI control, where editing the control returns an updated set of prompt options. In our study, the renderer was limited to generating radio buttons, checkboxes, and free-text areas, selected by the AI when generating options. Our approach can be extended to generate any UI elements that can be expressed using a declarative schema and converted into a GUI control. The serializer takes a set of prompt options and returns a textual prompt refinement that can be included in a prompt.

To generate prompt options, we prompt an LLM to return JSON conforming to a TypeScript schema (prompt detailed in Appendix~\ref{appendix:OptionModuleDefinition}). LLMs are highly capable at generating JSON and TypeScript code, and during the study we did not have to validate the result. To make the system more robust, structured output decoding~\cite{StructuredOutputs} can additionally be used to ensure the output conforms to a schema. The LLM streams the JSON and we use an incremental JSON parser to create and render prompt options as they are generated.

We define an \emph{options module} to be a component that takes input, or \emph{grounding}, invokes an LLM with the grounding, and returns a set of prompt options. Figure~\ref{fig:prompt_options} illustrates how the options module and associated components relate. An options module can implement either static or dynamic prompt middleware, depending on the grounding information and when the module is invoked. We now show how prompt options are generated and used in our tool.

\subsubsection{Inline options}
Figure~\ref{fig:dynamicFlow} presents the information flow in the \DYNAMIC system, and we describe how a user interacts with inline prompt options:
\begin{figure}
    \centering
  \includegraphics[width=0.85\textwidth]{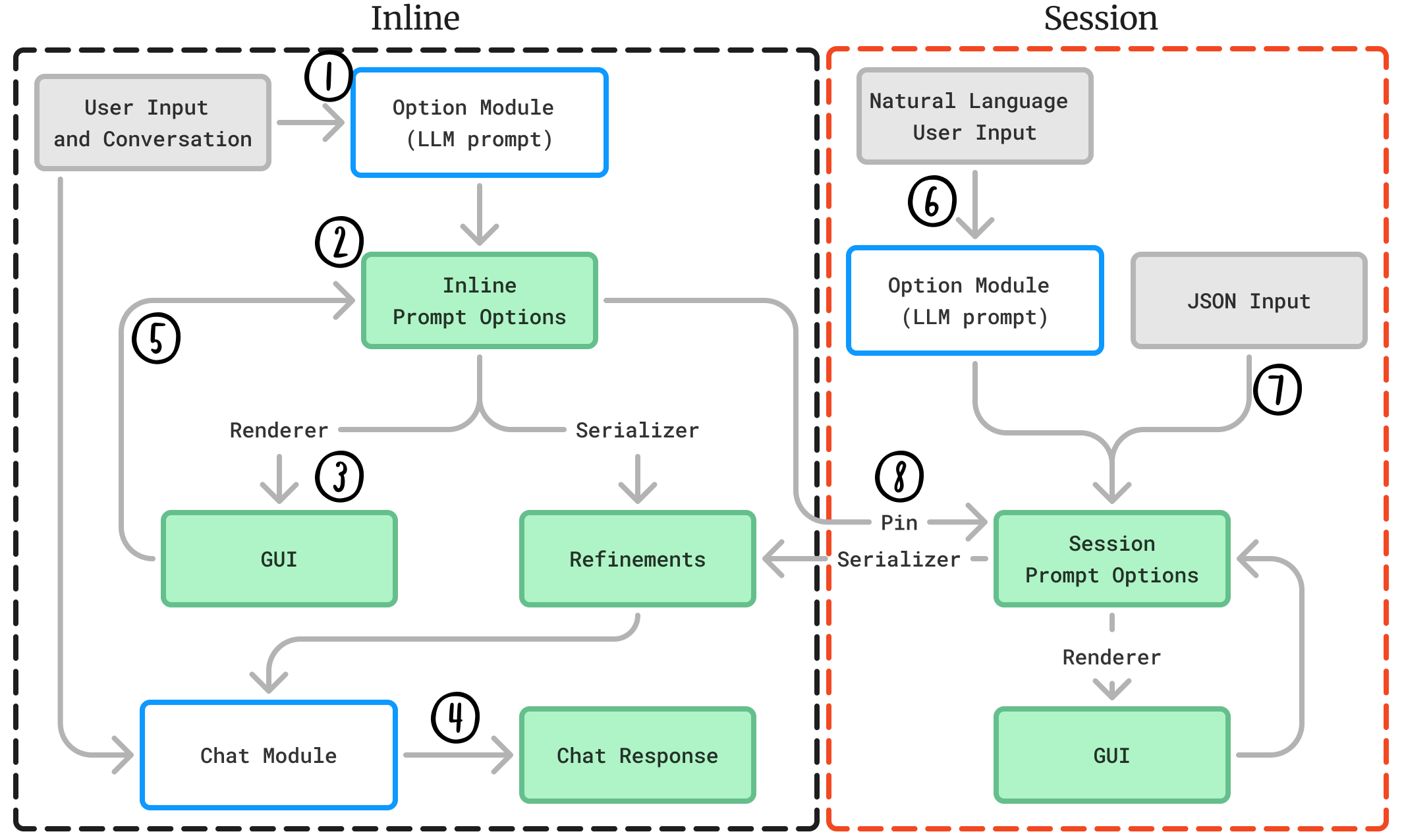}
  \caption{\DYNAMIC overall system flow. \emph{Inline:} (1) The \OPTIONMODULE takes the user's prompt input and conversation history. (2) The \OPTIONMODULE returns a set of prompt options with initial values based on the prompt. (3) Prompt options are rendered inline using a rendering engine. (4) The \CHATMODULE uses these refinements as grounding, along with the user's prompt input and conversation history, to generate a chat response. (5) User can adjust GUI controls which updates the refinements and re-invokes the \CHATMODULE to regenerate the current response. \emph{Session:} Users can directly invoke the \OPTIONMODULE to generate Session options through natural language prompting within the Session control panel. (7) Users can also directly add and modify options manually through JSON input in the Session control panel. (8) Users can pin \emph{Inline} options to the \emph{Session} by clicking the pin icon on each inline option, which applies the inline option to follow-up queries.}
  \Description{}
  \label{fig:dynamicFlow}
\end{figure}
\begin{enumerate}
    \item Given a user input, the conversation history, user input, and session-options are used as grounding to invoke the dynamic options module.
    \item The dynamic options module returns a set of prompt options, including initial values for the options.
    \item The prompt options are rendered inline using our rendering engine. The initial response does not require user interaction because the prompt options include initial values used by the serializer to generate refinements.
    \item These refinements are used as grounding, along with the input prompt and conversation history, to generate a response. 
    \item If the user adjusts a GUI control, the prompt options are updated, and the chat module is re-invoked to regenerate the current response with the new refinements in place.
\end{enumerate}

\subsubsection{Session options}
Session options are options that apply to every prompt and AI response in a given session with the \DYNAMIC system. 
Session options allow users to control what options are applied to every AI response, such as ``use bullet points'' which might apply to many prompts as a user preference.
Users have several ways to generate session options, which we describe with respect to Figure~\ref{fig:dynamicFlow}:
\begin{enumerate}\addtocounter{enumi}{5}
    \item Users can generate session options through a natural language utterance that is passed to an options module. The utterance can directly request an option, for example: ``I want to control expertise''. Alternatively, the utterance can be vague, or task driven, for example: `I am a novice programmer trying to learn''.
    \item Users can add or directly modify the JSON representing the session prompt options. By directly exposing the session options as JSON, they can be saved and reused in later sessions with the AI by copying the JSON code under the Chat controls UI and then pasting specific options into a new session as desired.
    \item Users can ``pin'' inline options that are generated by the system, which allows them to save inline options as session options. This action moves the option from the inline prompt options into the session prompt options and allows it to apply to all successive prompts within the session. For example, if the user is interacting with the AI to solve programming tasks, and the \DYNAMIC system generates ``Programming Language'' as an option. The user can then select their language of choice and then pin the option to the Session options so that the AI is given the context of what programming language the user is working in.
\end{enumerate}

The user can generate session options at any point in their workflow. They could generate session options before prompting the AI or generate new session options once the AI responds and ideas for new Session controls arise. In method (6) for generating session options, existing session options are used along with the user's utterance to generate new options.

Generation of session options in all cases relies on the user to directly request or define options. 
We do not recommend potentially useful session options nor generate them based on the user's prompt to the \CHATMODULE (these affordances might be useful in an improved \DYNAMIC system). 

\subsubsection{Generating chat responses}
\label{sec:DesignImpl:ChatModule}
The \CHATMODULE is a simple chatbot that interprets the conversation history and user message, similar to implementations like ChatGPT (prompt detailed in Appendix~\ref{appendix:ChatModuleDefinition}). Unlike these systems, the \CHATMODULE additionally receives prompt refinements from both the session and inline prompt options. The \CHATMODULE takes these inputs and generates a response which is then streamed to the user. The system is fully reactive: whenever a new session option is generated, or the user selects a different prompt option using a GUI, the latest AI response is automatically regenerated.

In this paper we apply the \DYNAMIC approach to a chat interface common to LLM interactions like Copilot, ChatGPT, and Gemini. However, the system is modular and can be used to help steer any LLM interface through the generation of options and refinements that are included in a prompt (see Section~\ref{sec:Discussion}).

\subsection{\STATICFULL}
\label{sec:DesignImpl:Static}
\begin{figure}
    \centering
  \includegraphics[width=0.65\textwidth]{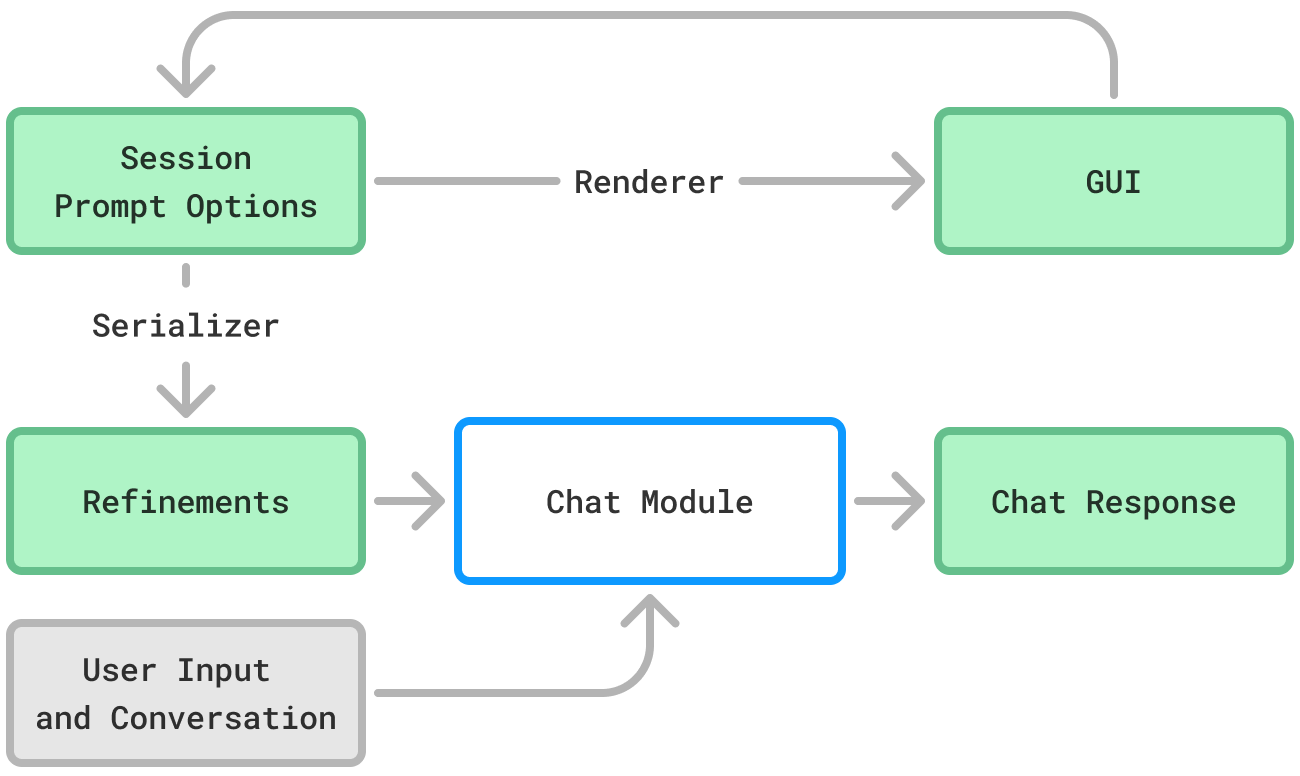}
  \caption{\STATIC model.}
  \Description{}
  \label{fig:StaticModel}
\end{figure}
\begin{figure}
    \centering
  \includegraphics[width=0.8\textwidth]{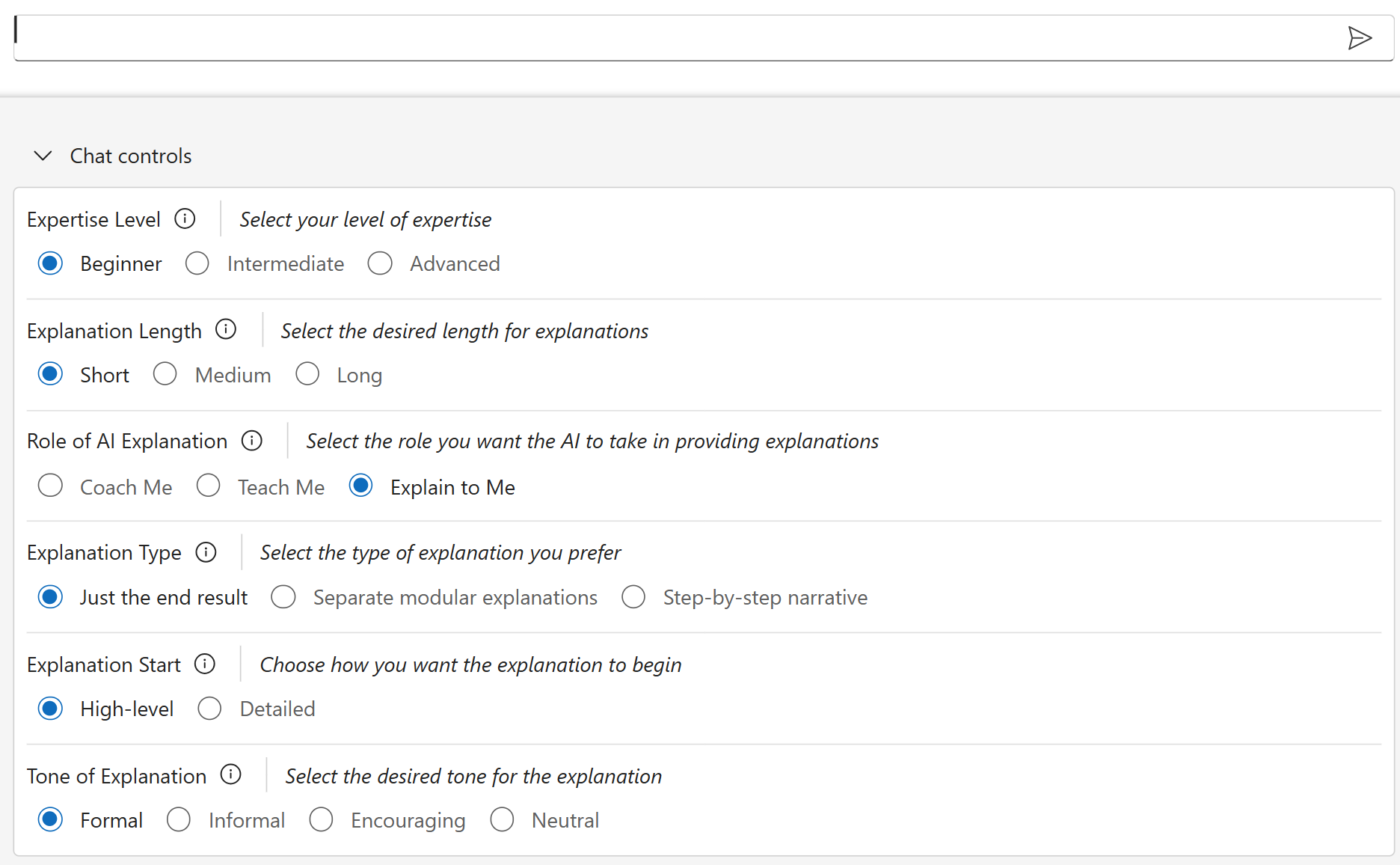}
  \caption{\STATIC options available during the study.}
  \Description{}
  \label{fig:StaticUX}
\end{figure}

In order to explore the usefulness of pre-selected controls and compare the experience to our \DYNAMIC approach, we also developed an alternative system called \STATIC (Figure~\ref{fig:StaticModel}) which has a few differences.
For \STATIC we used the same ChatModule but disabled the generation of inline and session options. 
Instead, \STATIC provides a list of pre-selected controls (detailed in Appendix~\ref{appendix:StaticOptionDefinitions}) derived from the results of our formative study (Section~\ref{sec:Formative}) that covers broad controls that might be useful for any prompt the user gives relating to explanations (Figure~\ref{fig:StaticUX}). 
There were several options requested by formative study participants that we did not add to \STATIC, as they involved options that required \DYNAMIC generation or interpretation or that were too specific to apply to the tasks more generally (e.g., \emph{``Talk to me like I'm a data scientist''}).
As with the options in \DYNAMIC, the \STATIC options will regenerate the current AI response on change which add option information and selection to the user's prompt.
The \STATIC system can be seen as adhering to design goal \textbf{D.1} (Section~\ref{sec:Formative:Results}) as it provides control over AI responses but does not adhere to \textbf{D.2} as the controls do not dynamically change to the task or user.

\section{Evaluation: In-lab comparative user study}
\label{sec:Methods}
To compare the impact of dynamic and static control on user needs for controlling AI responses in comprehension tasks, we conducted an in-lab comparative user study. This study informs the design of future prompt middleware systems for steering AI responses by addressing the following research questions:
\begin{enumerate}[label=\fcolorbox{black}{white!20}{RQ\arabic*},leftmargin=1.0cm, itemsep=-0ex]
    \item What are user preferences for control of AI responses in comprehension tasks?
    \item Which form of control is perceived as most effective for controlling AI responses? 
    \item How does each form of control impact user mental load? 
    \item What are the trade-offs of \DYNAMIC and \STATIC prompt middleware for controlling AI responses?
    
\end{enumerate}

We hypothesized that a \DYNAMIC approach would provide users more flexibility and control over a \STATIC approach, and be more preferred by users. However, we also predicted that there might be trade-offs for this increased control due to introducing complexity through new options that users would have to consider. 
On the other hand, the simplicity of \STATIC for controlling AI output may possibly feel more familiar to users (e.g., like adjusting settings of a desktop application), which could provide effective control without needing to use AI to generate dynamic elements for options. 
Beyond comparing \DYNAMIC and \STATIC approaches, we wanted to discover how to best design interactions for empowering users to get the AI responses they needed, without the tedium and effort of prompt engineering. 
Basically, we wanted to discover how to allow users to simply ask their question in \emph{real} natural language (that is, without requiring structuring prompts through prompt engineering for effective control of responses), and alleviate some of the challenges of prompting models, by providing simple UI interactions that allows users to steer the AI response.

\subsection{Participants}
\label{sec:Methods:Participants}
We recruited 16 participants (8 men, 8 women, 0 non-binary/other) via email from a list of generative-AI users who had previously signaled interest in participating in user studies. 
13 participants reported that they regularly used one or more GenAI tools (e.g., Copilot, ChatGPT, etc.) and three reported occasional use.
Since we chose formula and code comprehension tasks for our study, we also collected participant experience with spreadsheet formulas and programming. Most participants signaled basic use of spreadsheet formulas (12) with 3 having more advanced usage of spreadsheet formulas. 6 participants reported that they had never programmed, 4 having learned enough programming for small, infrequent tasks, and the remaining 6 participants being moderately (5) or highly (1) experienced in programming.
Finally, participants were located on various continents, with 9 in North America, 3 in Europe, 2 in Asia, 1 in Africa, and 1 in Australia. Participants were compensated USD \$50 or local currency equivalent for their time.

\subsection{Tasks}
\label{sec:Methods:Tasks}
Participants completed 6 tasks, organized into 2 task-sets, A and B (Figure~\ref{fig:tasks}). Each task-set involved 3 common tasks done with GenAI assistance: code explanation, complex topic understanding, and skill learning. The goal of each task was to interact with the system to craft an explanation based on each task prompt that helped the participant understand each task and made them confident that they could answer questions about the task with the explanation.

\begin{figure}
\textbf{Task Set A}
\begin{enumerate}[label=\fcolorbox{black}{red!20}{A\arabic*},leftmargin=1.0cm, itemsep=-0ex]
\item Explain the formula: 
\begin{lstlisting}
=INDEX(B2:E10, MATCH(G1, A2:A10, 0), MATCH(H1, B1:E1, 0))
\end{lstlisting}
\item Summarize and explain the following text: \texttt{[five paragraphs on impact of AI]}.

\item \textit{How do I take my data and find out if two things correlate with each other?}
\end{enumerate}
\noindent\rule{.85\linewidth}{0.4pt}

\textbf{Task Set B}
\begin{enumerate}[label=\fcolorbox{black}{green!20}{B\arabic*},leftmargin=1.0cm, itemsep=-0ex]
\item Explain the following Python code:
\begin{lstlisting}[language=Python]
import pandas as pd}
df = pd.read\_csv('data.csv')}
df['Total'] = df['Quantity'] * df['Price']}
result = df.groupby('Category')['Total'].sum().reset\_index()}
result = result.sort\_values(by='Total', ascending=False)}
\end{lstlisting}
\item Summarize and explain the following text: \texttt{[five paragraphs on impact of climate change]}.

\item \textit{How do I visualize my sales data for my website?}

\end{enumerate}
\caption{Task-sets A and B}
\label{fig:tasks}
\end{figure}

\subsection{Protocol}
\label{sec:Methods:Protocol}
We considered our two conditions for this within-subjects user study, \DYNAMIC and \STATIC (described in Section~\ref{sec:DesignImpl:Dynamic} and Section~\ref{sec:DesignImpl:Static} respectively). 
Every other element and interaction within the system was identical between conditions. 
We chose a within-subjects protocol so that participants could reflect on the similarities, differences, and trade-offs of using a \DYNAMIC versus a \STATIC. 
We did not compare \DYNAMIC or \STATIC to a `baseline' ChatGPT or similar system as our formative study results showed that users strongly desired control of AI-generated content, which we suspect would be replicated in the comparative tool study. 
Instead, we directly compare \DYNAMIC against \STATIC to better understand the effectiveness of different approaches of control for users, and the trade-offs of each. Each user study took approximately 90 minutes.

\begin{figure}
    \centering
  \includegraphics[width=0.95\textwidth]{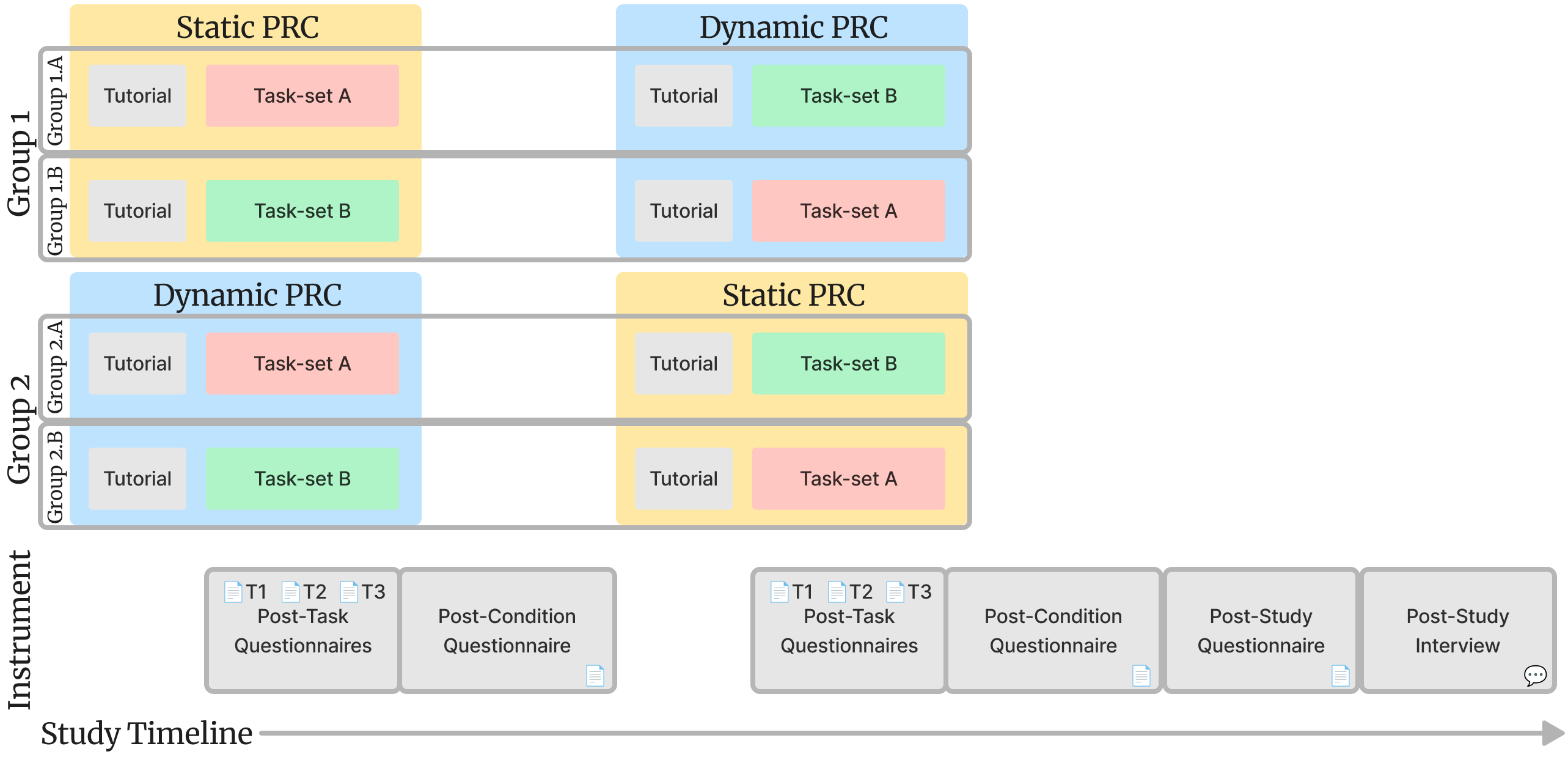}
  \caption{Overall design of the study.}
  \Description{}
  \label{fig:TaskBalance}
\end{figure}

Participants were assigned \DYNAMIC and \STATIC conditions and A and B task-sets through a counterbalanced design, such that half the participants received the \DYNAMIC condition first, and the other half received the \STATIC condition first. 
Within each of these condition-first groups, each task-set was balanced such that half of each condition saw the A task-set first, and the other half received the B task-set first. 
Therefore, there were four equal groups of participants during the study (see Figure~\ref{fig:TaskBalance}).

To mitigate learning and novelty effects, participants were first given a tutorial based on their first tool condition (either \DYNAMIC or \STATIC). 
Participants were shown how to interact with the system and the study facilitator walked them through example tasks like writing a hello world program in Python and planning a trip to London. 
In the \DYNAMIC condition, both tasks showed how the system generated ``inline'' options upon submission of a prompt, how to ``pin'' an inline option to session options, and how to delete an option.
Within the hello world tutorial task, participants were also shown how \DYNAMIC can generate options on command through the ``Chat controls'' panel by inputting \emph{``I am a novice programmer trying to learn''} and pressing ``Add controls'', which generated several new options within the session controls area.
In both conditions and in both tutorial tasks, participants were shown how selection from the available options regenerated the response based on their selection. 
Participants in both conditions were shown how to inspect the various informative tooltips on each option.
After the participant felt comfortable using the system, they were presented with their first task-set (either A or B).  

Participants then completed three tasks with their first condition. 
Participants were given 7 minutes per task to interact with the system to craft an explanation that they found useful. 
During each task, participants either modify option selection within the system, create new options if in the \DYNAMIC condition, or follow-up with the chat-module as necessary. 
At the end of 7 minutes, or if the participant was satisfied with the AI-generated response, the participant was asked three post-task questions (on a 7-point scale) which had the rate how much they believe the \emph{options} helped them get an explanation that helps them understand the task, and rate how confident they would be to answer questions about the task with the explanation in hand (reported in Section~\ref{sec:Results:ParticipantRatedTaskPerformance}, Appendix~\ref{appendix:PostTaskQ}).
After a task-set was completed within a condition, participants  filled out a questionnaire adapted from the NASA Task Load Index \cite{HART1988TLX} and the adaptations done related research \cite{Vaithilingam2024DynaVis} (reported in Section~\ref{sec:Results:ParticipantPerformance}, Appendix~\ref{appendix:PostConditionQ}).

Next, participants were given a second tutorial that introduced them to the other condition. 
As before, participants were shown how to interact with the system with the same example tasks shown in the first condition. 
Once the participant felt comfortable using the system, they moved on to completing their second task-set, like the first set of tasks, and completed post-task questionnaires. 
After these three tasks were completed, we again performed the post-condition questionnaire. 

Participants then completed a final post-study questionnaire that compared the two conditions on dimensions of system preference and ease of use (reported in  Sections~\ref{sec:Results:ParticipantPrefs} and ~\ref{sec:Results:MentalLoad}, Appendix~\ref{appendix:PostStudyQ}). 
This final questionnaire also elicited participant needs around control of AI responses, the importance of control of and learning from AI responses, and how helpful each condition of the system was (reported in Section~\ref{sec:Results:ParticipantAffectOnControl}). 
We performed and report Mann-Whitney U Tests to determine statistical significance between conditions.

During each task, participants were asked to think aloud while interacting with each system relating to their goals for crafting good responses with the AI, the options available to them to do so, and any other feedback and reactions they had while using the system. 
After all questionnaires were completed, researchers collected participant feedback about the user experience of each tool and their control needs through a semi-structured interview (Appendix~\ref{appendix:InterviewQuestions}). 
These interviews focused on the aspects of control and the systems the participants saw during the study. 
Questions prompted participants to think about other features that would provide control over AI, what other specific types of UI controls might be useful to generate, the importance of control of AI explanations, system preference, and reflecting on how their AI workflows might change with the tool. 
Participant utterances during the tasks were also analyzed as specific feedback for each condition. 

Transcripts of the user study were automatically generated, reviewed, and corrected as necessary by researchers, who then extracted participant utterances and an initial qualitative analysis using iterative open coding \cite{Vaismoradi2013OpenCoding} to identify representative themes, and thematic analysis \cite{Braun2006ThematicAnalysis} to condense and group utterances into related themes. 
Researchers then performed a second pass through negotiated-agreement \cite{McDonald2019InterRater, saldana2021coding} to further organize these themes to report successes and barriers for using \DYNAMIC and \STATIC prompt middleware, and the design implications for future \DYNAMIC systems. 
These final themes are reported in Section~\ref{sec:Results:Qual}.

\section{User Study Results}
\label{sec:Results}
\subsection{What are user preferences for control of AI responses in comprehension tasks? [RQ1]}
\label{sec:Results:ParticipantPrefs}

\subsubsection{Overall Comparison Between Conditions}
\label{sec:Results:ParticipantPreferenceForSystem}
\begin{figure}
    \centering
  \includegraphics[width=0.85\textwidth]{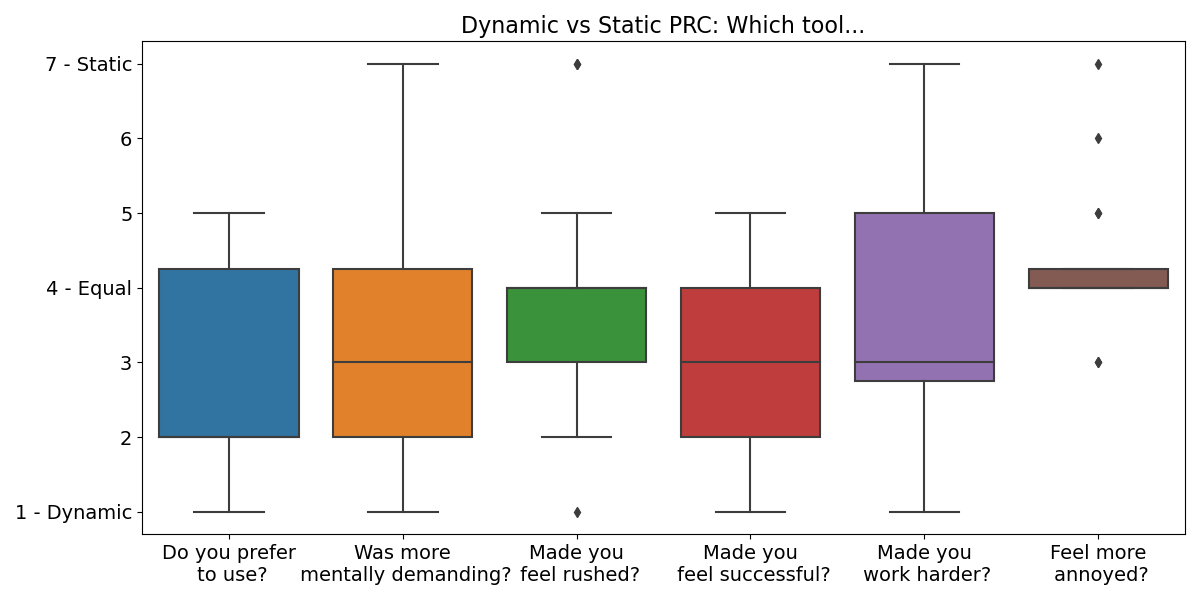}
  \caption{Overall direct comparison of conditions for participant tool preference, mental demand, feelings of being rushed, successful, level of effort, and feelings of annoyance.}
  \Description{Overall direct comparison of conditions for participant tool preference, mental demand, feelings of being rushed, successful, level of effort, and feelings of annoyance.}
  \label{fig:ConditionComparison}
\end{figure}

After both conditions were complete, participants answered a questionnaire that directly compared the \DYNAMIC and \STATIC conditions, where scores closer to 1 represent a stronger preference for \DYNAMIC, scores closer to 7 represent a stronger preference for \STATIC, and a score of 4 representing that the two systems were equal for the question (Figure~\ref{fig:ConditionComparison}). 

Participants reported that \DYNAMIC was slightly more demanding to communicate with (median=3.0, mean=3.31) and made participants work harder to accomplish their level of performance (median=3.0, mean=3.56). Participants thought the two tools were equal when considering which tool made them feel more rushed during the task (median=4.0, mean=3.88) and which tool made them feel more insecure, discouraged, irritated, stressed, and annoyed (median=4.0, mean=4.31). 
However, despite participants feeling like they had to work and think slightly harder while using \DYNAMIC to control AI explanation generation, the affordances that \DYNAMIC brought for controlling AI responses during each task led to participants responding that they preferred to use \DYNAMIC over \STATIC (median=2.0, mean=2.81) and reported that they felt slightly more successful at completing the tasks with \DYNAMIC over \STATIC (median=3.0, mean=3.0). 




\subsubsection{Participant Reported Preferences Around Control}
\label{sec:Results:ParticipantAffectOnControl}
\begin{figure}
    \centering
  \includegraphics[width=0.95\textwidth]{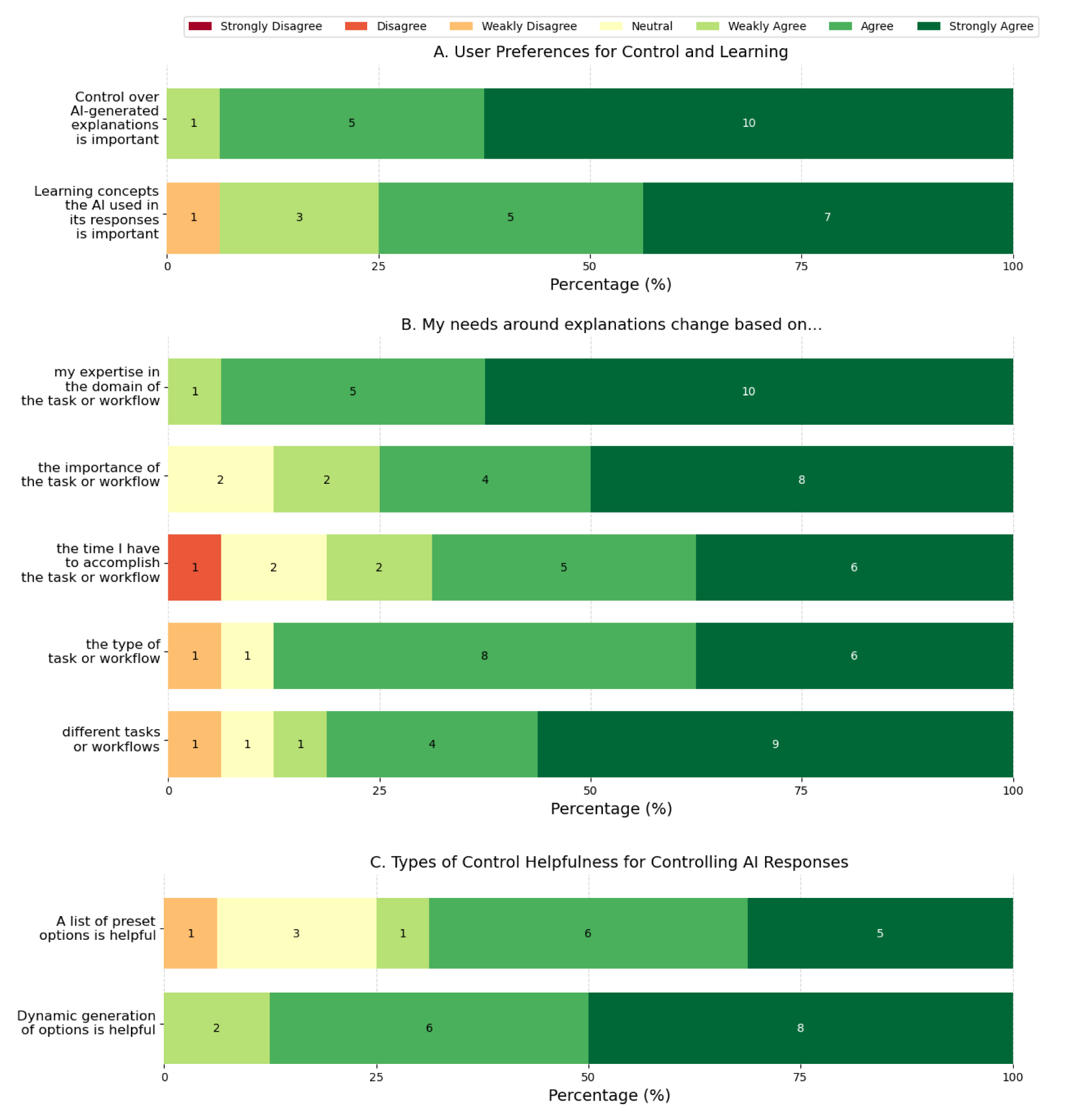}
  \caption{Participant reported preferences around control from Strongly Disagree to Strongly Agree. Sub-graph A. captures participant-rated importance of control of AI-generated explanations and of learning the concepts used by the AI. Sub-graph B. captures the elasticity of participant needs around AI-generated explanations based on several different factors. Sub-graph C. captures how helpful participants found each approach, \STATIC and \DYNAMIC, for controlling AI responses.}
  \Description{}
  \label{fig:CombinedPlots}
\end{figure}

After the completion of the study, participants completed a questionnaire on several dimensions of control of AI explanations to provide their preferences around control, reflect on the variability of their needs for AI-explanations based on various factors, and how helpful they find \DYNAMIC and \STATIC approaches for controlling AI-responses (Figure~\ref{fig:CombinedPlots}). 

Ratings of the importance of control over AI-generated explanations and of learning concepts, as motivation for improving the experience of explanation generation, (Figure~\ref{fig:CombinedPlots} A) showed that participants ``Strongly Agreed'' that control over AI-generated explanations was important to them (median=7.0, mean=6.56), but only ``Agreed'' that learning the concepts the AI used in its response was important (median=6.0, mean=6.06). 


Participant signaled that their needs around AI explanations changed frequently based on variability in their workflows (Figure~\ref{fig:CombinedPlots} B). In order of highest agreement rating: participants ``Strongly Agreed'' that their needs changed based on their expertise in the task or workflow (median=7.0, mean=6.56) and between tasks and workflows (median=7.0, mean=6.19). Participants rated between ``Agree'' and ``Strongly Agree'' that their needs change based on the importance of the task or workflow (median=6.5, mean=6.13). Finally, participants ``Agreed'' that their needs change based on the time they have to accomplish the task (median=6.0, mean=5.75) and on the type of task (median=6.0, mean=6.06).


Finally, participants rated the helpfulness of \DYNAMIC and \STATIC options around controlling AI-responses (Figure~\ref{fig:CombinedPlots} C). Participants more highly rated the helpfulness of \DYNAMIC over \STATIC for controlling AI responses (\DYNAMIC median=6.5 (between Agree and Strongly Agree), mean=6.38 vs \STATIC median=6.0 (Agree), mean=5.69). We elaborate on these findings by reporting themes found from our semi-structured interviews with participants in Section~\ref{sec:Results:Qual}.



\subsection{Which form of control is perceived as most effective for controlling AI responses? [RQ2]}
\label{sec:Results:ParticipantPerformance}
\subsubsection{\DYNAMIC vs \STATIC Effectiveness}
\label{sec:Results:ComparedEffectiveness}

\begin{figure}
    \centering
  \includegraphics[width=0.9\textwidth]{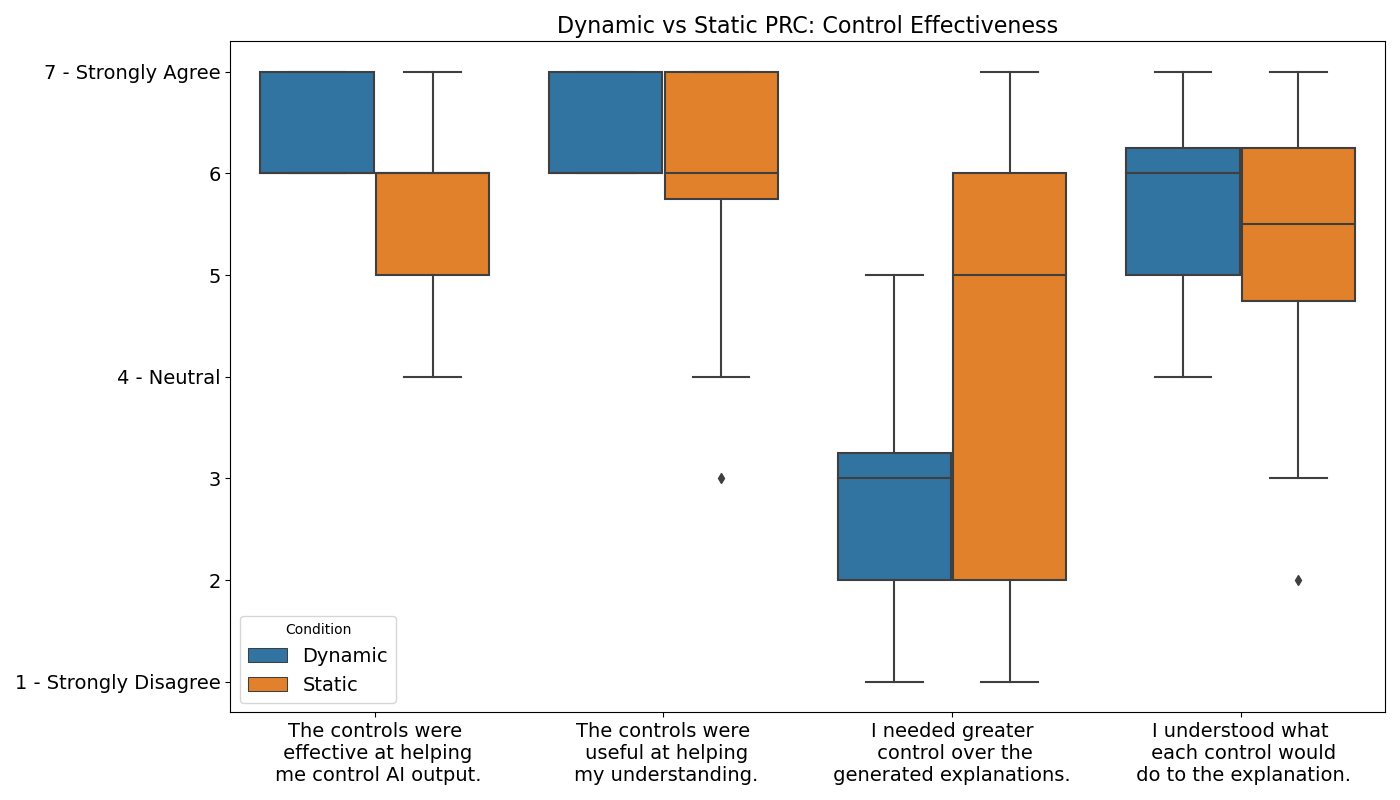}
  \caption{Participant rated effectiveness of each condition in the control of AI responses for the comprehension tasks they saw.}
  \Description{}
  \label{fig:ConditionEffectiveness}
\end{figure}

Participant reported on tool effectiveness by condition, on a 7-point Likert scale from Strongly Disagree (1) to Strongly Agree (7) (Figure~\ref{fig:ConditionEffectiveness}), and found the controls generated in \DYNAMIC to be significantly more effective at helping control AI output and explanations over \STATIC (\DYNAMIC median=6.0 (Agree), mean=6.44 vs \STATIC median=6.0 (Agree), mean=5.81; U: 73.5, P-value: 0.0245). 

We also found a significant difference between participant need for more control over the AI explanations generated during the tasks, where participants ``Weakly Disagreed'' that they needed more control with \DYNAMIC versus ``Weakly Agreeing'' in the \STATIC tool (\DYNAMIC median=3.0, mean=2.75 vs \STATIC median=5.0, mean=4.44; U: 190.5, P-value: 0.0174).

However, there was not a significant difference between conditions for how useful the controls were at helping participants better understand the concepts used in each task (\DYNAMIC median=6.0 (Agree), mean=6.44) vs \STATIC median=6.0 (Agree), mean=5.88), nor was there a difference for participant understanding of the impact that each control would have on the generation of an explanation (\DYNAMIC median=6.0 (Agree), mean=6.44 vs \STATIC median=6.0 (Agree), mean=5.31).

Participants rated the ease of use of each tool during the task and rated how well the AI understood their intent when selecting options.
and ``Agreed'' that it was easy to complete the tasks with both \DYNAMIC and \STATIC tools (\DYNAMIC median=6.0, mean=6.0 vs \STATIC median=6.0, mean=5.75). Participants also thought the AI in the \DYNAMIC tool understood their intent and made the right edits slightly more than in the \STATIC tool (\DYNAMIC median=5.5 (between Weakly Agree and Agree), mean=5.69 vs \STATIC median=5.0 (Weakly Agree), mean=5.5). However, we did not find a significant difference between conditions with either question.
Participants also reported similar success levels for completing their tasksets in both conditions on a scale from Perfect (1) to Failure (7), which was statistically insignificant (\DYNAMIC median=2.0, mean=2.19 vs \STATIC median=2.0, mean=2.5).

\subsubsection{Participant-Rated Task Performance}
\label{sec:Results:ParticipantRatedTaskPerformance}
\begin{figure}
    \centering
  \includegraphics[width=0.9\textwidth]{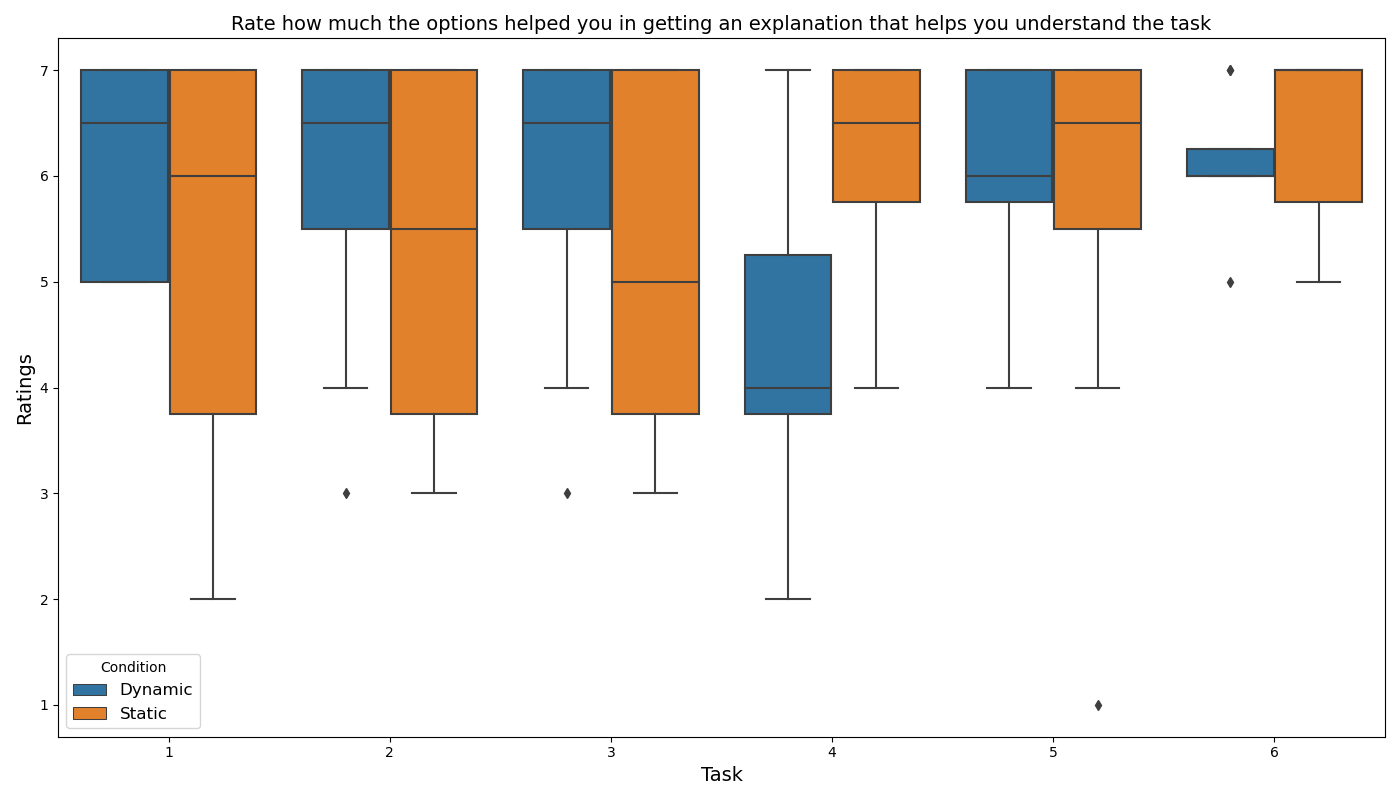}
  \caption{Measurement of how effective the options were in each condition at helping the participant control the response and assist the participant in understanding the task (higher is better).}
  \Description{}
  \label{fig:TasksRatingHelped}
\end{figure}
After each task was completed (either due to the participant signaling they were finished or after 7 minutes were up), participants rated from 1 to 7 on how much they believed the options helped in getting an explanation that helped them understand the task (Figure~\ref{fig:TasksRatingHelped}).
Only Task 4 (a code explanation task that involved explaining a Python code snippet) was found to have a significant difference between conditions with \DYNAMIC performing worse than \STATIC (U: 12.0, P-value: 0.0355). This potentially means that the options provided in \STATIC are an effective set of refinements for code comprehension tasks, but the system instructions for generating potential options through \DYNAMIC needs improvement to better support code explanations for users.

Similarly, as a measure of how useful the final result would be in accomplishing the task of understanding the content in the explanation, participants reported how confident they were that they could answer questions about the task with the explanation they generated in hand. However, we did not find any significant differences between conditions for each task in participant confidence levels.


\subsection{How does each form of control impact user mental load?  [RQ3]}
\label{sec:Results:MentalLoad}
\begin{figure}
    \centering
  \includegraphics[width=0.9\textwidth]{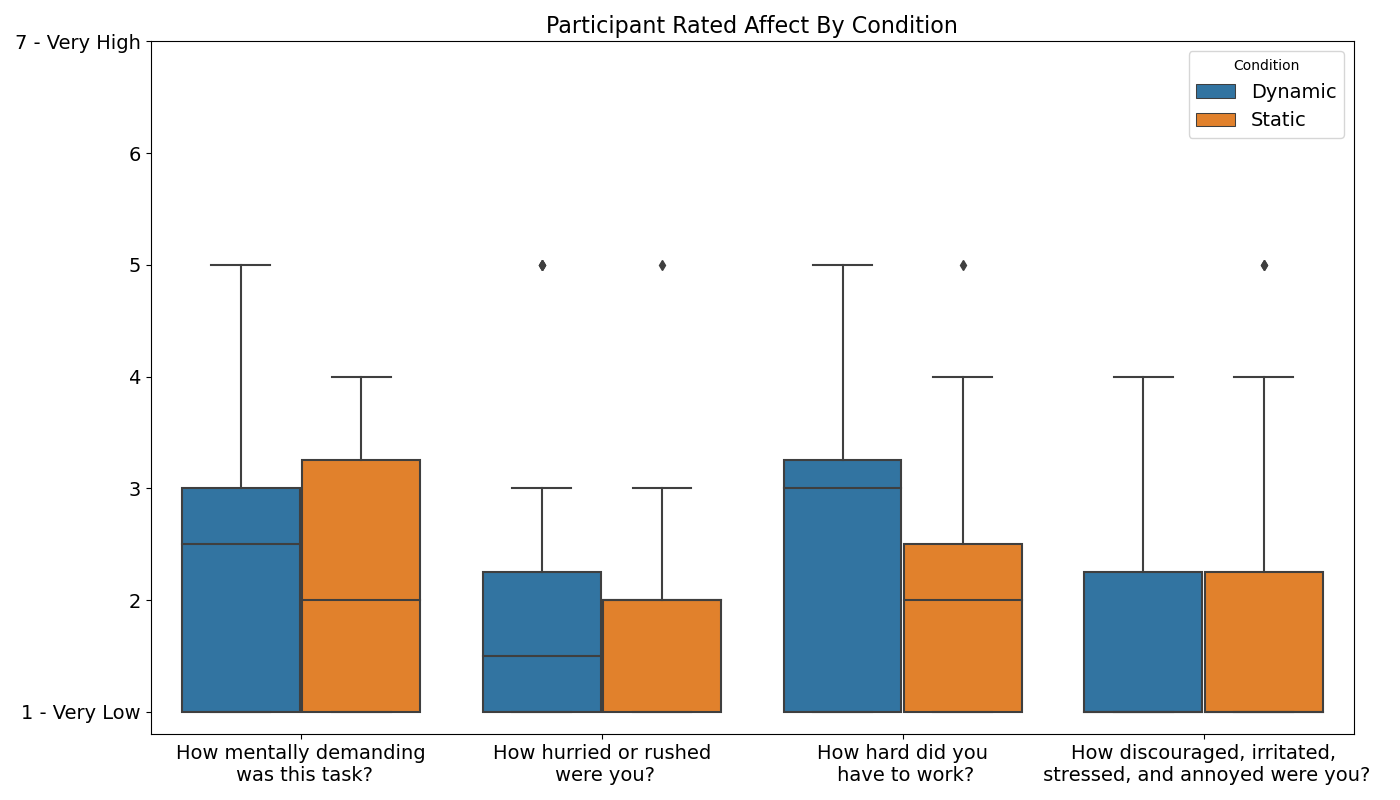}
  \caption{Participant rated mental load for each condition, rated after the completion of a task set.}
  \Description{}
  \label{fig:DemandTLX}
\end{figure}

Participants also reflected on the mental load felt while completing their task-sets within each condition from Very Low (1) to Very High (7) (Figure~\ref{fig:DemandTLX}). Participants reported that the tasks were slightly more mentally demanding in the \DYNAMIC condition (\DYNAMIC median=2.5, mean=2.44 vs \STATIC median=2.0, mean=2.25), felt slightly less rushed with \DYNAMIC (\DYNAMIC median=1.5, mean=2.13 vs \STATIC median=2.0, mean=1.94), felt like they had to work harder with \DYNAMIC (\DYNAMIC median=3.0, mean=2.69 vs \STATIC median=2.0, mean=2.13), and felt the same level of annoyance in both conditions (\DYNAMIC median=1.0, mean=1.81 vs \STATIC median=1.0, mean=1.88). However, we did not find a significant difference for any of these four categories of mental demand. 

We note that the results of this questionnaire report that participants felt slightly less rushed with the \DYNAMIC condition, the results for comparing conditions directly against each other (Figure~\ref{fig:ConditionComparison}) found an equal level of feeling rushed (though the mean was slightly towards \DYNAMIC). Although the two results show slightly different patterns, the findings presented here and in Figure~\ref{fig:DemandTLX} are not significant, and thus might be attributed to noise.

\subsection{Benefits and trade-offs of \DYNAMIC and \STATIC prompt middleware [RQ4]}
\label{sec:Results:Qual}

While completing the tasks in both conditions and during the post-study interview participants reflected on the impact of the dynamic and static prompt middleware they experienced, and the trade-offs of using these approaches while controlling AI responses. 
During interviews with participants and while using each version of the system, participants directly compared \DYNAMIC with \STATIC and reflected on the impact each system would have on their daily workflows with AI.

\subsubsection{\DYNAMIC lowers barriers to providing context to the AI}
\label{sec:Results:DynamicLowersContextBarrier}
Current participant workflows for prompt engineering was seen as tedious (P4), inconsistently effective (P4, P5), time-consuming (P5), and a current barrier participants had with using GenAI that new control affordances, like that of the \DYNAMIC system, could help address. Participants found it difficult to \emph{``even have the words for the context''} they wanted to convey in their prompts as there was a noticeable gap in expertise needed between prompting and prompt engineering (P4).
Providing task and user context within prompts was seen as critical for obtaining responses that aligned with user needs (P2). 

\DYNAMIC was described as helpful for assisting the users in providing much of the context necessary for the AI to effectively respond to a user's needs for a task and eliminate much of their struggles due to forming prompts (P2, P4, P8, P15). For example:  
\begin{quote}
\emph{``I have to do so much [work] around my prompt to give it the context to make it specific, point it to the tone to do, and a lot of times I'm repeating within those categories of types of prompts, I'm repeating those same things. I felt like dynamic would really help me shortcut what I'm doing manually today. (P15)}
\end{quote}

During the study, the inline options generated by the \DYNAMIC system replaced much of the need for follow-up prompts to the system in order to provide more context or correcting the initial prompt to refine the AI's response (P1, P3, P9, P16). 
In their current workflows with AI, participants described the back-and-forth between user and AI as having a negative impact on their productivity but said that \DYNAMIC would increase their performance and efficiency of their day-to-day workflows with AI since it lowers the need for forming follow-up prompts to the AI (P3, P4, P10, P16). 

\subsubsection{\DYNAMIC improves the perception of control over the AI}
\label{sec:Results:DynamicProvidesControl}
As reported in Section~\ref{sec:Results:ComparedEffectiveness} and \ref{sec:Results:ParticipantAffectOnControl}, participants felt that \DYNAMIC gave more effective control over the \STATIC system and their current prompting strategies (P4, P6, P8). 
Without the control provided by \DYNAMIC, one participant felt: 
\begin{quote}
    \emph{``restricted by the current Copilot [because I] don't have these options, I'm trying to modify the prompt to get the answer out and the AI just goes out of control after some time, so I have to refresh and start again and change my prompts.'' (P12)}
\end{quote}

The greater control afforded through generated options was seen as useful for receiving AI responses that fit the personal preferences of the user (P14), which participants felt was made possible through \DYNAMIC's afforded flexibility (P7), greater precision in defining the task (P8), better predictability of AI responses (P6), and greater adaptation to user needs (P13).
Participants felt that \DYNAMIC gave them greater control over the AI. This control allowed participants to directly define their response needs to the AI (P4) and guide the AI to a useful response that aligned with their preferences (P15). 
A few participants even likened the \DYNAMIC approach to programming and stated that the ability to modify the JSON to directly edit and fine-tune the options gave them more control over the AI (P6, P7). 
Finally, P10 saw control as a way to provide autonomy while working with AI: 
\begin{quote}
    \emph{``I think of this [AI] as my personal assistant[...] I want it to be positive that the AI is just a tool you use to kind of build up on some idea that you have. So, I think it's really important to be able to [have] control.''}
\end{quote}

\subsubsection{\DYNAMIC provides guidance for AI steering interactions}
\label{sec:Results:DynamicProvidesGuidance}
Participants saw the \DYNAMIC system as a guide in interacting with AI by providing helpful, and relevant, options (P1, P2, P4, P6, P7, P9, P14). For example, P7 saw the generated options as a form of debugging the AI's response since it \emph{``gives you some ideas around if you change [the response] in this way then you might get better answers.''} 
Participants felt that \DYNAMIC guided them towards better responses (P7, P10, P11), giving alternatives of what they might do to solve their problem (P7). 
\DYNAMIC options did this by helping participants review potential refinements to their prompt (P1, P4, P9) and give greater structure to their AI interactions through generated options (P14).
P11 said it even helped them learn potential strategies for future prompts, saying that \DYNAMIC was \emph{``giving me more of a prompt learning experience, and I'm getting out of [the AI] what I want. And actually, it's a better response.''}  

Participants saw \DYNAMIC as valuable for providing AI responses that had better detail and results than other methods like prompting or \STATIC gave (P1, P8, P11-13, P15), while requiring less focus on writing the prompt (P11), though this came at a cost of requiring slightly more effort than using the \STATIC (P12). 
By providing an interface that guided participants to explicitly provide their preferences, \DYNAMIC was seen as more efficient for helping steer the AI. P1 explains:
\begin{quote}
    \emph{``It's [\DYNAMIC is] all bumper rails at the bowling alley for getting to where you need to go, and you're trying to direct the ball that is the conversation you're having with this interface. And having more options to do that, I think, allows you to get there quicker.''} 
\end{quote}

\subsubsection{\DYNAMIC enables greater exploration of and reflection on tasks}
\label{sec:Results:DynamicProvidesReflection}

\DYNAMIC enabled our participants greater exploration of the AI's affordances (P1) and helped them better reflect on their tasks (P9). 
The inline options generated by \DYNAMIC were perceived by participants as a reflection of the user's own assumptions placed in the prompt that they could then correct to steer the response which also revealed alternatives to considerations they were making (P9, P10).
Other participants felt that the \DYNAMIC options provided a scaffold for their thought process on what they wanted to focus on for their explanation (P5, P7, P10, P11, P15) which was seen as helpful for users who are leveraging AI to complete tasks in an unfamiliar domain (P15, P16).
The \DYNAMIC options were helpful for getting users unstuck by providing alternatives (P8), and as a form of brainstorming with the AI (P16).
P1 said that \DYNAMIC helped them understand the possibilities of the AI model, since it generated options that \emph{``were things that I would maybe assume the system couldn't handle or that it wouldn't be able to do.''}  
Participants felt that having control through generated options enabled greater understanding of a task as it allowed users to craft responses to their personal preferences and learning needs (P6, P8). 

\subsubsection{\DYNAMIC and \STATIC have mixed effects on perceived cognitive effort}
\label{sec:Results:STATIC}
\paragraph{On ease of use}
Many participants thought that \DYNAMIC made their interactions with AI easier, by scaffolding their interactions programmatically (P6), giving users control over option generation through natural language (P4, P14), and by providing reuse of useful options for later use (P15). 
Several participants thought \DYNAMIC would even cause their AI usage to increase (P2, P4, P10, P13) since it lowered the effort needed form prompts (P4) and reduced time spent providing context to the AI (P2), which without prompt middleware has caused frustration and even abandonment of GenAI sessions (P10). 
Having access to the \DYNAMIC controls increased participant confidence in working with GenAI (P3), made it easier to ideate on goals for interacting with GenAI (P5), lowered the cognitive burden of working with GenAI (P9), and helped them better understand how to use GenAI (P5).

Despite participants' preferences towards using \DYNAMIC, \STATIC was also found to be helpful in steering AI-generated responses during the study but had several downsides.
Participants described a trade-off between the ease of use of the \STATIC system and the increased control and flexibility of the \DYNAMIC system (P1), as \STATIC options in the study were useful for many refinements that commonly needed when interacting with AI (P1, P2, P7 P11), similar to settings of a software application that are not frequently changed (P1). 

A few participants thought \STATIC allowed them to more easily define the kind of response they wanted before creating a prompt for the AI (P1, P12). For example, P1 said \STATIC... 
\begin{quote}
\emph{``allowed me to think a bit more before putting prompts in about what explanation am I looking for? What are my needs for this kind of task? Because it's hard to know what you don't know and what you want to ask, and having it there beforehand helped narrow that down.'' (P1)} 
\end{quote}

While \DYNAMIC also allows you to request options from the AI before giving providing any prompts, it is possible that the provided \STATIC options are simpler to think about (as the options only require the user to select their preference) vs the blank slate that is generating the \DYNAMIC Session chat controls, which likely contains similar barriers users have with a new AI chat session.

\paragraph{On mental load} 
However, some participants felt that there was a trade-off between having more control with the \DYNAMIC system over \STATIC as it provides a more complex interaction. 
Some participants thought that \DYNAMIC would require the user to think more as they had to consider each generated option and whether they needed to generate options that the AI did not recommend (P1, P10). 
\DYNAMIC was seen as more complicated as the options varied throughout the tasks (i.e., were dynamic) and \emph{``required more activation [mental] energy''} (P10), which made gave an edge to \STATIC in usability while under time-pressure (P13, P16) and for straight-forward tasks (P16).
This increased mental load was not always seen as a negative for all participants. For example, P13 \emph{``did not see `mentally demanding' as it is a negative thing''} since it encouraged them to think more about their task.

\paragraph{On perceived level of control}
Despite the usefulness of the \STATIC, many participants still wanted more control (P3, P4, P6-8, P10, P15 and Section~\ref{sec:Results:ParticipantPreferenceForSystem}). \DYNAMIC was seen as necessary for providing specific context to the AI (P2), as some \STATIC options had limited impact in several cases compared those in the \DYNAMIC (P2), including not impacting the response enough to improve their understanding of the topic (P4), not being specific enough to afford effective control of the AI (P10), and not appropriately affecting the response based on the selection (P3, P6). 

Participants felt that \STATIC did not give them the same control that \DYNAMIC did over the assumptions the \CHATMODULE made (P14), which meant some participants needed to craft follow-up prompts, which included asking clarifying questions to form better understanding (P6, P16), or iterating with the AI to get greater detail in the response (P8).
\STATIC also had barriers to understanding what options to select based on the task, which was exacerbated due to lack of expertise. For example, P3 felt that they were \emph{``not able to accurately identify how different configurations may help understand the concept''} due to not having much domain knowledge in the task, which also caused struggle in forming follow-up prompts.

Other participants wanted to craft new options that were unavailable in \STATIC that they thought would have been useful to have (P6, P7) and wanted options that took the context of the prompt or their data so they could focus the response on certain themes (P4, P10), which are available in \DYNAMIC but not \STATIC. 


\subsubsection{Mismatches between how AI interprets options and how users want options applied}
\label{sec:Results:UserAIMismatches}
While some participants felt that the control provided by the options increased the predictability of AI responses (Section~\ref{sec:Results:DynamicProvidesControl}), one of the most common challenges participants had during the study was understanding how the options they selected might be applied by the AI.  
This challenge manifested when there was a mismatch between the participant's expectations for how the option would be, or should be, applied by the AI, and how the option was actually reflected in the response to the prompt, which could potentially cause frustration in users (P10).
This gap between user expectations and AI application can come from user not understanding the potential impact of an option (P2-6, P8),
being uncertain on whether the option was applied by the AI (P1), 
or due to the prompt refinements produced by the \DYNAMIC system or by developers in the \STATIC system being insufficient to make an impact to the response (P1, P4, P8, P10, P12, P13, P15, P16). 

Uncertainty on how an option might be applied by the AI caused participants frustration when trying to understand differences between option selections (e.g., \emph{``Teach me''} vs other AI roles found in the \STATIC options (P2)) and when option selections modified parts of the response that participants did not expect (e.g., the expertise option modifying the number of examples given (P3)).

During the post-study interview, some participants wanted more fine-grained UI elements to be generated (e.g., sliders for controlling the count of example (P7)), but many participants discussed how these less discrete options might further exacerbate the barrier to expected impact between user and AI. Participants thought this was because it would be challenging to for them to understand the difference between finer-grained selections (P1, P4, P5, P9, P15), which would increase the time it took to select a refinement (P5), increase their mental burden compared to discrete radio buttons (P4, P9) as it was difficult to predict how the AI's response would be impacted by subtle differences (P4).

For other options, it was difficult to know if they were applied due to their subjective nature (P1). One way the system helped participants understand if an option was used by the AI was when the response explicitly mentioned an option selection:
\begin{quote}
\emph{``I noticed that there was often little paragraphs at the end explaining what had been done and, I don't know if I'm naively, just assuming that because that's the options I've specified and that they would have the impact that they would have in that response.'' (P1)}
\end{quote}

Even when participants felt that an option clearly had impact, some options were not applied to the level of effect expected or needed (P8, P10). For example, P8 thought that by selecting the ``beginner'' and ``coach me'' options would better explain the concept in terms they could understand as they did not have experience with programming, and expressed dissatisfaction that the \STATIC options did not modify the response enough to assist in their understanding of the code.

Other mismatches occurred for when modifying length impacted what systems the AI recommended in the response (P12, P13), when modifying expertise changed the topic focus of a text summarization in task 2 (P15), and with an option around complexity changing the length of the examples in a response (P16).
While working with the system, participants even began forming their own mental models of how the AI would apply the options. For example, one participant initially believed that the order of the options (from top to bottom) reflected on where the AI would apply the options in the response (P1).

Another method participants used to better understand how the AI would apply options, and to find the most effective options, was to quickly select each option and see how it was reflected in the response since the systems quickly regenerated a new response on selection (P4, P5, P8).
This gave participants a greater understanding of how the model interpreted the option for their prompt, how it impacted the response, and discover when certain options were less effective. 
For example, while P6 thought they understood the majority of the \STATIC options, they \emph{``didn't find a real difference between beginner and intermediate (expertise)''} until they modified response length to be greater than ``Short'' and observed that \emph{``some options nullify other options that you put in.''} 
It is possible that these methods were sufficient in addressing the barrier around understanding the impact of the options on the AI's responses for participants, as reflected in Figure~\ref{fig:ConditionEffectiveness}. 

\section{Discussion}
\label{sec:Discussion}

\subsection{Design Implications for Dynamic Prompt Middleware}
\label{sec:Discussion:DesignImplications}
From study participant reflections, we derive three implications for improving the design of \DYNAMIC: provide control over how options are applied, leverage relevant context, and allow direct manipulation of options.

\paragraph{\DYNAMIC should provide control of how and where the AI applies options}
Though participants believed the control afforded by \DYNAMIC was useful and made them more effective (Sections~\ref{sec:Results:DynamicLowersContextBarrier} - \ref{sec:Results:STATIC}), participants felt that more control over the application of options could help address the barriers that remain (described in Sections~\ref{sec:Results:STATIC} and \ref{sec:Results:UserAIMismatches}).

\begin{figure}
    \centering
\includegraphics[width=0.50\textwidth]{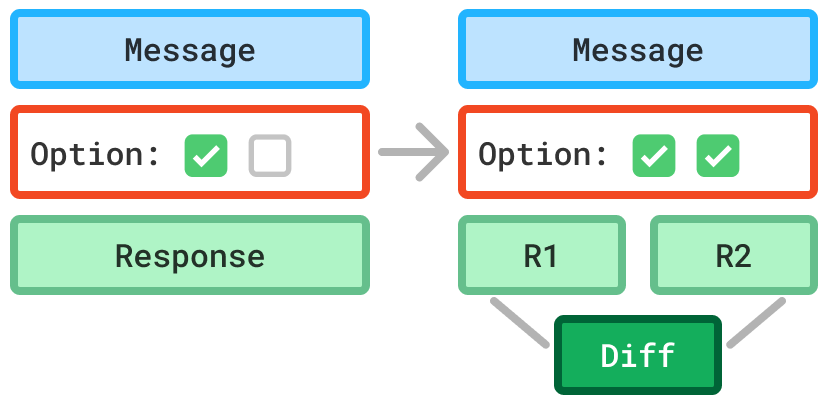}
  \caption{PRC exploration support can be improved by generating diffs of the various option selections for users.}
  \Description{}
  \label{fig:dyanmicDiff}
\end{figure}

Participants noted a need for greater transparency on how the AI interprets the options selected and prompt given to the AI (P1, P14, P15). 
This might be achieved through providing several potential examples of the structure and format of the response for selection (P3, P8)), supporting interventions within an AI response where users prompt the AI to adjust specific sections of a response by directly interacting with the response to communicate intent (P6, P10, P15), or even provide response diffs (similar to code diffs \cite{diff}) to allow comparisons between responses based on the selections made by the user to address barriers to user understanding around how the option selections impacted the response (P1). The current design of the \DYNAMIC system is capable of supporting such response branching and exploration (Figure~\ref{fig:dyanmicDiff}), but the addition of informative diffs would help address the barrier to user understanding of the impact of options on responses.

Another useful ability would be to allow the user to make targeted changes to the \textit{response}, so that users can keep results that they like, and refine what they do not through \DYNAMIC interactions. The user might wish to preserve some aspects of the structure of the response while refining the content (P14) which can be lost when modifying prompts.

Finally, participants were frustrated with how the AI applied subjective options (like those found in the \STATIC system around role of the AI). 
Therefore, generated refinements should be evaluated for subjectiveness and broken down into objective sub-refinements. This will better support users in steering the behavior of the AI in applying prompt refinements to the response, and increase the predictability of refinement impact. 

\paragraph{\DYNAMIC should gather and leverage relevant context for the user} 
While the current implementation of \DYNAMIC uses the user's prompt and currently selected options to generate options, participants wanted the system to leverage their past AI usage and other context or tool usage to inform the generation, and selection, of \DYNAMIC and provide greater context to the AI (related to Section~\ref{sec:Results:DynamicLowersContextBarrier}). 
For example, for code generation tasks systems could use previous AI code generation requests to infer what programming language to use (P10), or incorporate the files the user has open in their IDE. 

As with the previous need for understanding how AI interprets options, transparency was cited as a critical need for understanding and controlling what data the AI uses to form a response (P6, P7, P9, P12).
Forming user personas based on information like user-provided expertise level and job position (P2) and interacting with the user to let the user select what relevant data AI should use to form responses (P1, P11, P12, P15) were also seen useful features for greater specificity and personalization of responses.
This need might be fulfilled by explicit signals from the AI in the response on when data comes from an external source, or from the model itself, by highlighting each within a response (P6). 
Users might then select these highlights to steer the AI toward using sources and data that the users prefer (P4, P10). 

\begin{figure}
    \centering
\includegraphics[width=0.90\textwidth]{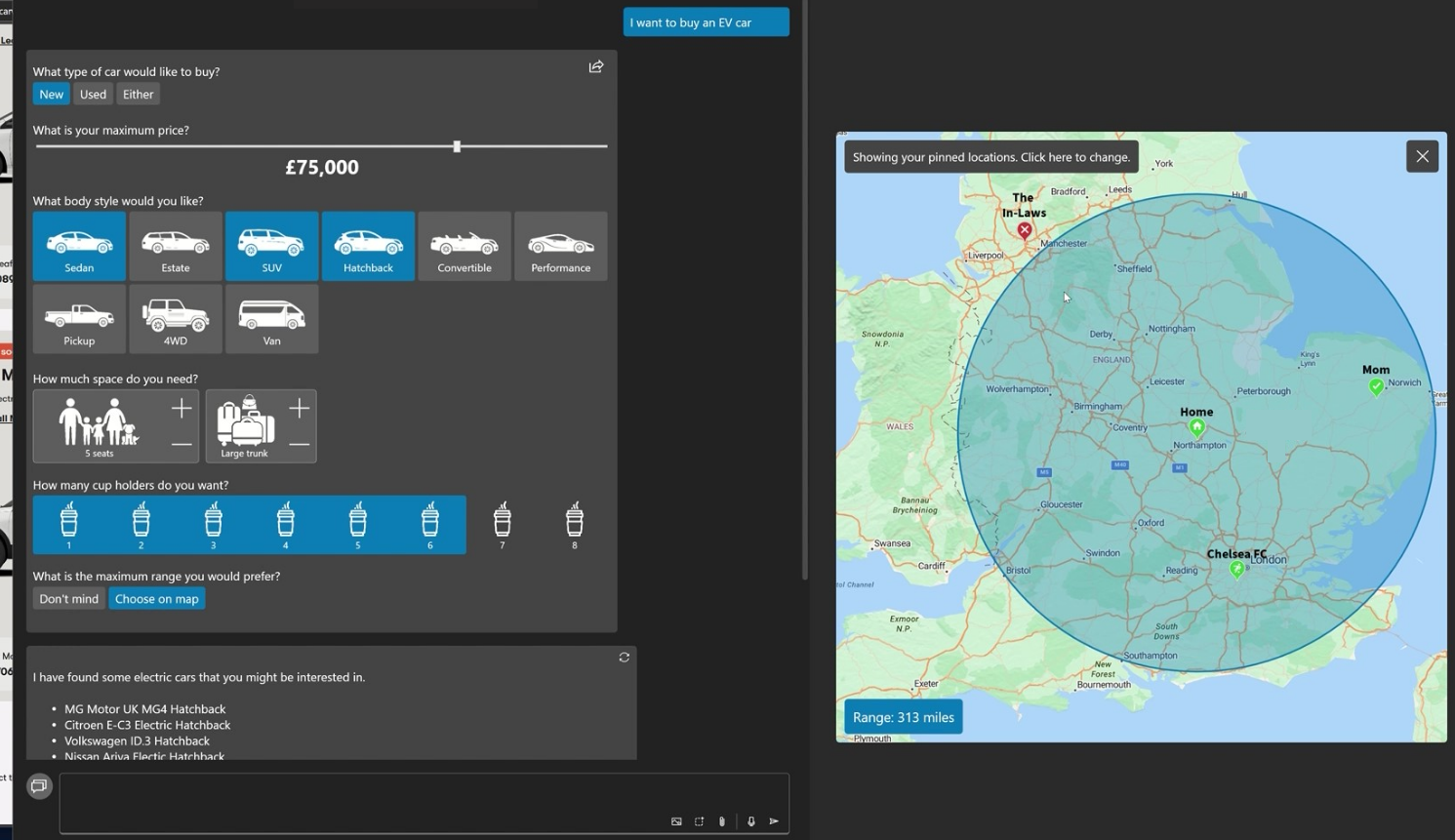}
  \caption{High fidelity prototype showing richer PRC modalities like interactive maps to choose EV range and visual representations of options to collect context for helping a user buying an EV car.}
  \Description{}
  \label{fig:richerModalities}
\end{figure}

\DYNAMIC systems might also generate richer PRC modalities, like interactive maps to reveal and collect geographic needs or visual representations of options to provide the user inspiration for what kind of context the AI is looking for, which might improve the process of context gathering from the user (Figure~\ref{fig:richerModalities}).
However, there must be a balance between the amount of context-gathering interactions that the user must attend to, so that information overload and fatigue in providing preferences to the \CHATMODULE is avoided. This issue can potentially be eased through PRC option reuse, similar to what session options in the \DYNAMIC and options in the \STATIC approaches provide.

\paragraph{\DYNAMIC should provide direct manipulation to update options} While it was uncommon for participants to want to modify the generated \DYNAMIC options during the study (as participants could either prompt the \OPTIONMODULE to regenerate the option with changes or modify the JSON representation), some felt the need to directly update options (e.g., click on an option to edit the text and regenerate the option based on the changed text) (P2, Section~\ref{sec:Results:DynamicProvidesControl}). 
The \DYNAMIC system can be improved to allow users to update parts of the UI (e.g., the option labels) to send these modifications to the \OPTIONMODULE as user intent. Based on this intent, the \OPTIONMODULE could regenerate the option to better suit the user's needs, supporting a form of response-by-example.

\subsection{Impact of \DYNAMIC on Cognition Beyond Prompt Control}
Our study showed that while \DYNAMIC's \OPTIONMODULE focused on generating options that help gather greater context and provide users' adaptable control of the AI explanations and addressed barriers to using GenAI found in previous research~\cite{drosos2024duck, YanIvie2024}, it also enabled greater reflection and exploration of comprehension tasks for users (Section~\ref{sec:Results:DynamicProvidesReflection}). 
This suggests that \DYNAMIC can potentially act as a tool for thought by assisting user thinking around their tasks. We see \DYNAMIC options as potential metacognitive interventions that can encourage self-awareness and reflection~\cite{tankelevitch2024metacognitive} during prompting and interacting with AI.
Building on \DYNAMIC's ability to scaffold the user's task by providing guidance on how to steer the AI and generating alternatives (Section~\ref{sec:Results:DynamicProvidesGuidance}), the system could be enhanced to also scaffold users' thought processes about their task or even challenge users' selections, encouraging them to reconsider their choices when relevant by acting as an ``AI provocateur''~\cite{sarkar2024aiprovocateur}, rather than just an AI prompt options assistant.
However, this direction creates tension between the original goal of providing users dynamic control over AI and in fostering critical thinking. Therefore, future research should explore when these thought interventions would be helpful, without reducing the autonomy and control afforded by Dynamic PRCs.

\subsection{Generative UI versus Programming Support for Generative AI}
Recently there has been much discussion of the potential for using GenAI to create entire user interfaces, personalized to the user and the task at hand. Poupyrev's vision talk for ACM UIST 2023 speaks of the ``ultimate interface'' \cite{poupyrev2023ultimate}, and the influential UX research consultancy Nielsen Norman Group refers to this as Generative UI \cite{moran2024generative}. While our approach relates to these superficially in that we also generate UI components dynamically, our view is that the automatic generation of user interfaces is better suited as a means for increasing user control over GenAI, rather than as an end in itself. The reason for this is that approaching the challenges of GenAI use as being similar to \emph{programming} the system for performing a task, rather than directly performing the task themselves, is more fundamentally aligned with how knowledge workflows shift in response to the introduction of GenAI \cite{sarkar2023aiknowledgework}. In other words, users tend not to seek customized interfaces within which they can perform their task, rather they seek better interfaces for eliciting and manipulating GenAI-produced artifacts. Indeed, this is the approach taken by the recent UIST workshops on \textit{Architecting Novel Interactions With Generative AI Models} \cite{InteractionXGenAI} and \textit{Dynamic Abstractions} \cite{dynamic_abstractions_uist24}. Petricek theorizes programming interfaces as being a series of ``substrates'' \cite{petricek_2022} -- modes of programming that trade off increasing levels of power and expressivity against increasing complexity and decreasing learnability (e.g., spreadsheet formulas are easy to learn, but limited in their expressivity; VBA macros are more complex, but allow greater programmability). GenAI has the potential to at least partly defuse some of these trade-offs, and bridge multiple substrates using a single interface (natural, or naturalistic language) \cite{sarkar2023eupgenai}, thus creating a smooth gradient of programmability for end users.

\subsection{Limitations}
One limitation of our study is that we did not compare \DYNAMIC and \STATIC to a baseline chat (e.g., ChatGPT, Copilot, or Gemini). Our formative survey results informed this decision as the vast majority of respondents (89.5\%) believed having flexible control over the AI's explanation to be important. As such, we focused on two potential approaches of prompt middleware to discover trade-offs and effectiveness in providing control to users. As we selected participants with GenAI experience, they were able to draw on their previous interactions with GenAI to compare \DYNAMIC and \STATIC to their current AI workflows during our interviews.

Our tasks involved comprehension of data and programming-centric scenarios. While we have used \DYNAMIC to generate refinements in a multitude of scenarios beyond these (e.g., ``Help me fix my leak'', ``Plan a workshop''), we did not evaluate \DYNAMIC and \STATIC potential for assisting users steering AI responses beyond comprehension tasks. Future research should consider how \DYNAMIC impacts user experience in the breadth of applications of GenAI and study how the effects of \DYNAMIC over longer periods of usage.

\DYNAMIC has the ability to pin, edit, and save the refinements generated. While some participants explored these features, we did not evaluate their usage or effectiveness for users to understand how \DYNAMIC might be reused between several tasks.

Finally, there may have been effects due to the system instructions of the \OPTIONMODULE and \CHATMODULE which might impact the effectiveness of the systems. We do not evaluate the impact of changing these instructions and iterated during development of the \DYNAMIC system until it reached a level of reliability required to evaluate in our study. 
One potential effect of not evaluating the effectiveness of our prompts may be the occurrence of the mismatch between human and AI barrier on how options should be applied (Section~\ref{sec:Results:UserAIMismatches}). 
It is possible that improvements to the system prompts in our \DYNAMIC system might improve the effectiveness of the controls generated and obtain greater alignment between user and AI.

\section{Conclusion}
\label{sec:Conclusion}
This study explored how to address the challenge of effective prompting to control and steer AI responses to comprehension tasks. 
Informed by a formative survey of GenAI users ($n=38$), we implemented two approaches of prompt middleware, \DYNAMICFULL and \STATICFULL to better understand the trade-offs between predictable prompting support and adaptable and contextualized support.
Our within-subjects study ($n=16$) found a preference for the \DYNAMIC approach they felt it afforded more control of the AI, lowered barriers to providing context, and encouraged greater exploration and reflection. 
This research contributes design implications for future \DYNAMIC systems that enhance user control of AI responses and suggests that dynamic prompt middleware can improve the user experience of generative AI workflows by affording greater control and guide users to a better AI response. 
However, barriers to reasoning about the effects of different generated controls on the final output still remain for these approaches, which calls for a greater collaboration between XAI systems and systems that afford greater control of AI-generated explanations.


\bibliographystyle{ACM-Reference-Format}
\bibliography{sample-base}

\appendix
\section{System Prompts}
\subsection{\OPTIONMODULE}
\label{appendix:OptionModuleDefinition}
\begin{lstlisting}[basicstyle=\scriptsize]
const instructions: string = `
# START INSTRUCTIONS: 
<role> You are an AI chat bot</role>
<task> Based on the conversation context, consider the important aspects and dimensions of the response, detail that 
should be included, or other options. Generate a set of options that would be useful for generating the response. 
</task>
<rules>
<rule> You must consider these instructions for generating a new UI element to control AI-generated explanations based 
conversation. </rule>
<rule> Consider options like level of detail, output types, learning objectives, and other factors that may be relevant 
to the user query that would impact a response. </rule>
<rule> An AI will use these controls to generate responses that are relevant to the user query and user preferences, 
so only generate controls that would be useful and relevant in guiding an AI in generating responses. </rule>
<rule> Consider the user query and provide a collection of multiple controls (between 3 and 5) and options (between 3
and 5) most suitable to the user's task, goal, workflow, and intent. </rule>
<rule> If the options do not conflict with each other, choose appearance of "multi-select-checkbox" to show checkboxes
to select multiple options. </rule>
<rule> If the options might conflict with each other, choose appearance of "single-select-radio" to show radio buttons
to select one option. </rule>
<rule> If the user provides a more open ended prompt (e.g., "I am a data scientist" or "I am a programmer"), respond
with multiple new relevant options (between 2 and 5 relevant options) that might include options the user would not
think of controlling (be creative). </rule>
<rule> If the user provides a request (e.g., "Explain this code"), respond with multiple controls (between 3 and 5
relevant options) that is most relevant to the user query and does not conflict with the current prompt or chosen
options. </rule>
<rule> If the user is specific on the type of control they want, respond with that option with 3-5 relevant options or
scales. </rule>
<rule> DO NOT GENERATE REDUNDANT options OR OptionControls (avoid generating similar labels or descriptions) that are
already present in the UI (listed below in <options></options>). </rule>
<rule> Try to generate the most diverse set of relevant options possible based on the user query and the current
options. </rule>
<rule> The generated options should also not conflict with the intent of the user (e.g., selection of Python libraries
when the user is asking about JavaScript code or has not signaled they are using Python). </rule>
<rule> First start with broad options, and then narrow down to more specific options based on the user query as more
options are added in <options></options>. 
-- For example: if the user asks about a specific feature, consider other potential related features or concepts a user
might ask about and generate options for those as well.
-- For example: if the user asks to perform a specific task, consider other potential related tasks a user might ask 
about and generate options for those as well. </rule>
<rule> ALWAYS CONSIDER THE INTENT OF THE USER, not just their prompt when generating new options. </rule>
<rule> Generate options that are relevant to the user's intent and what they are trying to achieve. </rule>
<rule> Consider each part of the user's potential workflow and generate relevant options for them to select 
from to guide the AI in generating responses and explanations.
-- For example: if the user asks to explain code, generate options that help them understand the code better and try to
predict options that are useful for follow-up queries, not just options that explain the specific piece of code 
in general.</rule>
<rule> Try to create a profile or persona of the user based on the options they have selected and use this when 
generating new options. </rule>
<rule> Predict what the user might ask next and generate options that would be useful for those queries based on
user intent, query, and current option selections (values). </rule>
</rules>
<format>

# !Given the following typescript schema, return a list with JSON object(s) that satisfies the type 'Option[]' 
based on the user query!
# !DO NOT RETURN ANYTHING ELSE BUT THIS LIST OF OPTIONS IN JSON FORMAT!
\`\`\`typescript
/** Option control */
interface OptionControl {
    type: "option";
    label: string; /** The label for the option */
    description: string; /** Short natural language description for the control */
    options: Record<string, string>; 
    /** Format: Record<label, description>. A dictionary of option labels to descriptions.
    Descriptions should include or mention the label for comprehension and represent 
    instructions to an LLM for generating explanations. */
    appearance: "single-select-radio" | "multi-select-checkbox";
    /** YOU MUST recommended the initial value in 'value' from options based on the user prompt and options.
     * Use the value of the selected option(s), not the key.
     * For example, if the option you are selecting is {"verbose": "Provide verbose detail", ...}, value should be the
     description ("Provide verbose detail"), not label ("verbose").
     * If \`appearance\` is "multi-select-checkbox", \`value\` is an Array of strings (string[]) */
    value: string | string[]; 
    reason: string; /** Reason for adding the control and how an AI might use it in a response
    (why is it important to consider the option added?) */
}

/** Text field option (when open-ended input from the user is better than discrete options)*/
interface TextControl {
    type: "text";
    label: string;
    description: string; /** Short natural language description for the control */
    value: string; /** YOU MUST recommended the initial value based on the user prompt and options */
    reason: string; /** Reason for adding the control and how an AI might use it in a response
    (why is it important to consider the option added?) */
}

type Option = OptionControl | TextControl
\`\`\`

# Example output for new controls based on the user prompt "I want explanations to match my expertise in 
data analysis":
[{
type: 'option',
appearance: 'single-select-radio',
label: 'Expertise level',
description: 'My expertise in data analysis',
options: { 'Novice' : 'I am a novice at data analysis', 
'Intermediate' : 'I have some experience in data analysis', 
'Expert' : 'I am an expert in data analysis'},
value: 'I am a novice at data analysis',
reason: 'Used to provide explanations that match your expertise in data analysis.'
},
{
type: 'option',
appearance: 'multi-select-checkbox',
label: 'Data Visualization Preferences',
description: 'Choose your preferred types of data visualization',
options: {'Graphs': 'Focus on graphical representations like charts and plots', 
'Tables': 'Prefer tabular data presentation', 
'Interactive': 'Include interactive visualizations for deeper insights'},
value: ['Focus on graphical representations like charts and plots', 'Prefer tabular data presentation'],
}],
reason: 'Used to determine which data visualization types will be considered in the generating a response.'
</format>

<conversation_history>
${currentContent}
</conversation_history>

<important>
If this content is the same as the user query, generate options based on the user query and what the user 
is attempting to do.
For example, if the user is trying to understand a complex concept, generate options that help the user 
understand or learn about the concept better.
</important>

# The existing options are:
<options>
${currentOptions}
</options>
#END INSTRUCTIONS
`;
\end{lstlisting}

\subsection{\CHATMODULE}
\label{appendix:ChatModuleDefinition}
\begin{lstlisting}[basicstyle=\scriptsize]
const startInstructions = `--- Instructions: You are a helpful assistant providing explanations 
to a user.
- You must consider the options selected (in "value") and user preferences for dictating the 
characteristics of the explanation you provide before generating any explanations in your response.
- If the options are missing or conflicting, use your best judgement to generate a response.
- If the options are irrelevant to the user's prompt, you should ignore them.
- If you use an option, be sure to reference it somewhere in the response if relevant.
- You may use Markdown formatting in your response.
#START SELECTED OPTIONS
<options>
`;
const endInstructions = `</options>
#END SELECTED OPTIONS
`;
\end{lstlisting}

\subsection{\STATIC Options}
\label{appendix:StaticOptionDefinitions}
\begin{lstlisting}[basicstyle=\scriptsize]
[{type: "option",
    label: "Expertise Level",
    description: "Select your level of expertise",
    options: {
      Beginner: "I am a beginner with limited knowledge",
      Intermediate: "I have a moderate level of expertise",
      Advanced: "I am highly knowledgeable and experienced",
    },
    appearance: "single-select-radio",
    value: "I am a beginner with limited knowledge",
    reason: "This control helps tailor the complexity of the explanations to match your understanding,
    ensuring that the information is accessible and useful for your level of expertise.",},
  {type: "option",
    label: "Explanation Length",
    description: "Select the desired length for explanations",
    options: {
      Short: "Provide concise, to-the-point explanations",
      Medium: "Provide moderately detailed explanations",
      Long: "Provide comprehensive, in-depth explanations",
    },
    appearance: "single-select-radio",
    value: "Provide concise, to-the-point explanations",
    reason: "This control allows you to specify the length of explanations to suit your preference and
    time constraints, ensuring that the information is delivered in a manner that is most useful to you.",},
   {type: "option",
    label: "Role of AI Explanation",
    description: "Select the role you want the AI to take in providing explanations",
    options: {
      "Coach Me": "Guide me through the process, providing tips and corrections as needed",
      "Teach Me": "Provide educational insights and foundational knowledge",
      "Explain to Me": "Clarify concepts or procedures directly without additional guidance or teaching",
    },
    appearance: "single-select-radio",
    value: "Clarify concepts or procedures directly without additional guidance or teaching",
    reason: "This control allows you to tailor the AI's approach to explanations according 
    to your learning preference or current needs, enhancing the effectiveness of the information provided.",},
   {type: "option",
    label: "Explanation Type",
    description: "Select the type of explanation you prefer",
    options: {
      "Just the end result": "Provide only the final outcome or answer",
      "Separate modular explanations": "Break down the explanation into distinct, modular parts",
      "Step-by-step narrative": "Provide a detailed, step-by-step narrative of the process",
    },
    appearance: "single-select-radio",
    value: "Provide only the final outcome or answer",
    reason: "This control allows you to specify how detailed and in what format you want the 
    explanation to be, which can help tailor the response to your understanding or needs.",},
    {type: "option",
    label: "Explanation Start",
    description: "Choose how you want the explanation to begin",
    options: {
      "High-level": "Start with a high-level overview of the topic",
      Detailed: "Begin with a detailed explanation of the topic",
    },
    appearance: "single-select-radio",
    value: "Start with a high-level overview of the topic",
    reason: "Allows you to control the initial depth of the explanation to match your 
    preference or current understanding level.",},
  {type: "option",
    label: "Tone of Explanation",
    description: "Select the desired tone for the explanation",
    options: {
      Formal: "Use a formal and professional tone",
      Informal: "Use a casual and conversational tone",
      Encouraging: "Use an encouraging and positive tone",
      Neutral: "Maintain a neutral and objective tone",
    },
    appearance: "single-select-radio",
    value: "Use a formal and professional tone",
    reason: "Allows the user to customize the tone of the explanation to match their 
    preference or the context in which they are using the information.",},
]
\end{lstlisting}

\section{Study Instruments}
\label{appendix:Instruments}
\subsection{Post-Task Questionnaire}
\label{appendix:PostTaskQ}
\paragraph{1. Rate how much the options helped you in getting an explanation that helps you understand the task (1-7 Lowest to Highest).}	
\paragraph{2. How confident are you that you can answer questions about this task with this explanation? (i.e., how much does this explanation help you?) (1-7 Lowest to Highest).}	
\paragraph{3. What other options do you think would be useful? [Free-Text]}

\subsection{Post-Condition Questionnaire}
\label{appendix:PostConditionQ}
\paragraph{1. Answer each statement with your level of agreement (From Strongly Disagree to Strongly Agree):}
\begin{enumerate}
    \item The controls were effective at helping me control AI output and explanations.
    \item The controls were useful at helping me better understand the concepts involved in the tasks.
    \item I needed greater control over the explanations generated during the tasks.
    \item I understood what each control would do to the explanation.
\end{enumerate}

\paragraph{2. Answer each statement with your level of agreement (From Strongly Disagree to Strongly Agree):}
\begin{enumerate}
    \item It was easy to complete the tasks using the tool provided.
    \item The AI understood my intent and made the right edits.
\end{enumerate}

\paragraph{3. Rate how successful you were in the tasks you saw (From Perfect to Failure):}
\begin{enumerate}
    \item How successful would you rate yourself in accomplishing this task?
\end{enumerate}

\paragraph{4. Rate your experience during the tasks (From Very Low to Very High):}
\begin{enumerate}
    \item How mentally demanding was this task with this tool?
    \item How hurried or rushed were you during this task?
    \item How hard did you have to work to accomplish your level of performance?
    \item How insecure, discouraged, irritated, stressed, and annoyed were you?
\end{enumerate}

\subsection{Post-Study Questionnaire}
\label{appendix:PostStudyQ}
\paragraph{1. Compare each tool (A and B) for the below statements and choose your preference (From 1 - A to 7 - B):}
\begin{enumerate}
    \item Which tool would you prefer to use?
    \item Which tool was more mentally demanding to communicate?
    \item Which tool made you feel hurried or rushed during the task?
    \item Which tool made you feel successful in accomplishing the task?
    \item For which tool did you work harder to accomplish your level of performance?
    \item Which tool made you feel more insecure, discouraged, irritated, stressed, and annoyed?
\end{enumerate}

\paragraph{2. Rate your agreement to the statements below (From Strongly Disagree to Strongly Agree):}
\begin{enumerate}
    \item Control over AI-generated explanations is important to me.
    \item Learning concepts the AI has used in its responses is important to me.
    \item My needs around AI explanations change based on the time I have to accomplish the task or workflow.
    \item My needs around AI explanations change based on the importance of the task or workflow.
    \item My needs around AI explanations change based on the type of task or workflow.
    \item My needs around AI explanations change based on my expertise in the domain of the task or workflow.
    \item My needs around AI explanations change between tasks or workflows.
    \item Dynamic generation of options is helpful for controlling AI responses
    \item A list of preset options is helpful for controlling AI responses
\end{enumerate}

\subsection{Semi-Structured Interview Questions}
\label{appendix:InterviewQuestions}
\begin{enumerate}
    \item \emph{What other features would help you express control over AI explanations?}
    \item \emph{What other types of controls would be useful to you?}
    \item \emph{Is control of AI explanations important to you, why?}
    \item \emph{Which system do you prefer? Why?}
    \item \emph{If you had access to the tools you saw today, how might your workflows with AI change?}
\end{enumerate}

\end{document}